\documentclass[bibyear]{aa}  

\usepackage{graphicx}
\usepackage{natbib}  
\usepackage{float}
\usepackage{color}
\usepackage{txfonts}
\usepackage{gensymb}
\usepackage{amsmath}

\def\src {PSR~B0943+10}

\begin{document} 

\newcommand{\pone}{P_1}
\newcommand{\ptwo}{P_2}
\newcommand{\pthr}{P_3}
\newcommand{\hpthr}{\hat{P}_3}
\newcommand{\hptwo}{\hat{P}_2}
\newcommand{\RPA}{R_\mathrm{PA}}
\newcommand{\rhoG}{\rho_{1\,\mathrm{GHz}}}
\newcommand{\cp}{cycles/$\pone$}
\newcommand{\thph}{\theta(\phi)}
\newcommand{\TC}{\Theta_\mathrm{C}}
\newcommand{\TLC}{\Theta_\mathrm{LC}}
\newcommand{\TTC}{\Theta_\mathrm{TC}}
\newcommand{\nulo}{\nu_\mathrm{lo}}
\newcommand{\nuhi}{\nu_\mathrm{hi}}

\title{\src: low-frequency study of subpulse periodicity in the Bright mode with LOFAR}

   \author{A.~V.~Bilous                     
          }

    \institute{Anton Pannekoek Institute for Astronomy, 
    University of Amsterdam, Science Park 904, 
    1098 XH Amsterdam, The Netherlands, 
    \email{A.Bilous@uva.nl}
         }

\abstract{We use broadband sensitive LOFAR observations in the 25--80\,MHz frequency range 
 to study the single-pulse emission properties of the mode-switching pulsar B0943+10. 
We review the derivation of magnetospheric geometry, originally based on low-frequency radio data, 
and show that the geometry is less strongly constrained than previously thought. This may be used to help 
explain the large fractional amplitudes of the observed thermal X-ray pulsations from the polar cap, which 
contradicted the almost aligned rotator model of PSR B0943+10. We analyse the properties of drifting subpulses 
in the Bright mode and report on the short-scale (minutes) variations of the drift period.
We searched for the periodic amplitude modulation of drifting subpulses, which is a vital argument for 
constraining several important system parameters: the degree of aliasing, the orientation of the line-of-sight 
vector with respect to magnetic and spin axes, the angular velocity of the carousel, and thus, the gradient of 
the accelerating potential in the polar gap. The periodic amplitude modulation was not detected, indicating that 
it may be a rare or narrow-band phenomenon. 
Based on our non-detection and review of the available literature, we chose to leave the aliasing order unconstrained 
and derived the number of sparks under different assumptions about the aliasing order and geometry angles. Contrary 
to the previous findings, we did not find a large (of the order of 10\%) gradual variation of the separation 
between subpulses throughout Bright mode. We speculate that this large variation of subpulse separation 
may be due to the incorrect accounting for the curvature of the line of sight within the on-pulse window.
Finally, we report on the frequency-dependent drift phase delay, which is similar to the delay reported previously 
for PSR B0809+74. We provide a quantitative explanation of the observed frequency-dependent drift phase delay within 
the carousel model.}

\keywords{pulsars -- individual sources (PSR~B0943+10, PSRJ~J0946+0951) -- telescopes (LOFAR)}

\titlerunning{Subpulse periodicity of \src\ during Bright mode below 100\,MHz}
\maketitle

\section{Introduction}
\label{sec:intro}

\src\ (originally named PP~0943, with PP standing for ``Pushchino pulsar'') is one of several dozen pulsars that 
have been discovered in the late 1960s, during the very first few years of pulsar astronomy 
\citep{Vitkevich1969}. It is a relatively old (characteristic age $\tau\approx5$\,Myr), non-recycled pulsar, 
mostly known for its two stable, recurring modes of electromagnetic emission. 

Once in a few hours, \src\ switches abruptly between so-called ``Bright'' and ``Quiet'' modes 
\citep[hereafter B and Q;][]{Suleymanova1984}. At radio frequencies, the pulsar is a few times brighter 
in B mode, and its single pulses form regular drifting patterns within the on-pulse longitude  window \citep{Taylor1971}. 
With the onset of Q mode, the drift disappears, the flux density drops, and the shape of the average profile changes
\citep{Suleymanova1984,Suleymanova1998}. Mode transitions also
affect the properties of the high-energy 
emission of the pulsar \citep{Hermsen2013}. In X-rays, the pulsar is $\sim2.4$ times \textup{brighter }in the 
radio-quiet mode, and the emission has a different spectral shape than in B mode. The X-ray emission from 
both modes has a pulsed thermal component that originates in a hot polar cap \citep{Mereghetti2016}.

The broadband nature of the mode-switching phenomenon may point to some global-scale magnetospheric 
transformation during the mode transition \citep{Timokhin2010}. However, the exact mechanism of the mode-switching remains unclear. 

Observing \src\ with the new generation of sensitive low-frequency radio telescopes with large fractional 
bandwidths may provide valuable input to solving the mode-switching puzzle. At frequencies below 200\,MHz, 
the average profile morphology starts to evolve rapidly in the
B and Q modes. Within the framework of 
radius-to-frequency mapping theory \citep{Cordes1978}, this means that below 200\,MHz, a greater range 
of emission heights can be observed at once. At the same time, the pulsar is quite bright in this frequency 
range, so that it is possible to track the changes in the average profile shape on timescales of a few minutes 
and also to observe the individual pulses directly. 

In our previous paper \citep[][ hereafter B14]{Bilous2014}, we used 21 hours of LOFAR observations (mainly 
in the 25--80\,MHz frequency range) to explore the frequency evolution in the average profile of \src\  and to follow the gradual changes in profile shape within several mode instances. We here 
explore the B-mode subpulse drift in the 25--80\,MHz frequency range based on the three observing sessions from 
B14. The subpulse drift is traditionally explained to occur through the influence of the gradient of the accelerating 
potential in the polar gap, which causes the fixed configuration of the emitting region to rotate slowly around the 
magnetic axis \citep[``carousel model'',][]{Ruderman1975,vanLeeuwen2012}. 
Thus, the drift properties are linked directly to conditions in the polar gap and provide a handle on the 
magnetospheric state in general. 

For \src, the properties of drifting single pulses were also used to establish the basic magnetospheric 
geometry: the angles between spin and magnetic axes, and the impact angle of the observer's line of sight 
(LOS). Based on the analysis of the frequency-dependent separation between B-mode average profile components  
and the brief presence of a 37-period modulation in the amplitudes of drifting subpulses, 
\citet[][hereafter DR01]{Deshpande2000} concluded that \src\ is an almost aligned rotator. Such a geometry is 
hard to reconcile with the large fractional amplitude of the polar cap X-ray pulsations that were found recently 
\citep{Mereghetti2016}. 

After a brief introduction of the observing setup in Sect.~\ref{sec:obs}, in Sect.~\ref{sec:geom} we 
review the derivation of the geometry in DR01 and test it against LOFAR observations. We outline the data processing and 
drift parameterisation in Sect.~\ref{sec:sps}. In Sects.~\ref{sec:Qsubp} and \ref{sec:Bsubp} we present the 
search for subpulse periodicity in both modes, confirm the lack of periodicity in Q mode, and describe 
the observed properties of the drift in B mode. In Sect.~\ref{sec:phtr} we attempt to constrain the number 
of sparks using so-called ``phase tracks'' (variations of immediate drift phase within the on-pulse window). 

Owing to the broadband nature of our observations, we were able to detect a so-called frequency-dependent 
phase delay, first discovered for PSR B0809+74 by \citet{Hassall2013}. In addition to variation within the on-pulse window, 
single pulses at different parts of the observing band have different drift phases at the same time. In 
Section~\ref{sec:fdphd} we show for the first time that the frequency-dependent phase delay of \src\ 
has a simple quantitative geometric explanation within the carousel model.
We give a short summary of our findings and conclude in Sect.~\ref{sec:summ}.

\section{Observations}
\label{sec:obs}

\begin{table}
\caption{Summary of observations.}             
\label{table:data}      
\centering                        
\begin{tabular}{lll}     
\hline\hline                 
\rule{0pt}{2ex} ObsID& Date &  Observing time per mode (hr)\\  
\hline                  
\rule{0pt}{2ex}L99010& 27 Feb 2013 &  0.3 Q $\rightarrow$ 3.6 B \\
L102418& 09 Mar 2013 &  2.0 Q $\rightarrow$ 2.0 B \\
L169237& 21 Aug 2013 &  2.0 B $\rightarrow$ 0.5 Q $\rightarrow$ 4.0 B $\rightarrow$ 1.4 Q  \\
\hline                                   
\end{tabular}
\end{table}

\src\ was observed with the low-band antennas (LBAs) of the LOFAR core stations at three epochs, with a total 
observing time of 15.8 hours. Raw complex voltages of two linear polarisations from each 
station were coherently summed, and the total intensity samples were recorded in a filterbank format.
We here work with the  uncalibrated total intensity signal. For 
the expected amounts of fractional linear and circular polarisations of \src\ \citep{Rankin2006b}, 
the contamination of Stokes $I$ by the leakage from the other components of the Stokes vector is expected to be small, 
of the order of few percent \citep{Kondratiev2016,Noutsos2015}.

The pulsar was detected in the frequency range of $25-80$\,MHz. A short 
summary of the observations is given in Table~\ref{table:data}, which lists the observation ID (ObsID), 
the date, and pulsar mode coverage for each session.  For details on the observing setup and initial 
data processing, we refer to B14.

\section{Magnetospheric geometry of \src}
\label{sec:geom}

\begin{figure}
   \centering
 \includegraphics[scale=0.8]{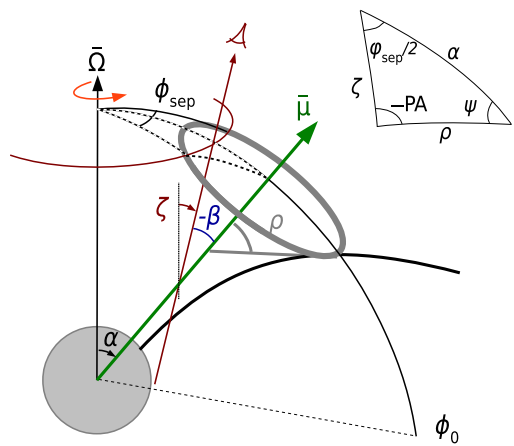}
 \caption[]{Sketch of basic magnetospheric geometry. The angular momentum $\bar{\Omega}$ of the pulsar is directed 
 vertically and the magnetic dipole vector $\bar{\mu}$ is inclined by the angle $\alpha$. The angle between the spin
 axis and the LOS is the viewing angle $\zeta$. The LOS vector stays fixed in space while the pulsar rotates around the 
 spin axis in counter-clockwise direction. The sketch shows the moment in time when the magnetic dipole and 
 LOS vectors lie on the same plane, corresponding to the fiducial rotational longitude, $\phi_0$. In this plane, 
 the angle between an observer and the magnetic axis is $\beta$. Following EW01, $\alpha$ and $\zeta$ are measured 
 clockwise from the direction of $\bar{\Omega}$, and $\beta$ is measured clockwise from the direction of 
 $\bar{\mu}$ (see text for more details). The thick black line shows an example of a dipolar field line, and the thick 
 grey line marks the emission cone with an opening angle, $\rho$. Two components of the observed 
 average profile are separated by $\phi_\mathrm{sep}$ in rotational longitude.  The dashed-line  spherical triangle 
 is shown in the inset, with PA being the measured position angle of the linearly polarised radiation, and 
 $\psi$ being the magnetic azimuth.  We note that $\beta$, $\zeta$, and $\rho$ are usually plotted shifted
to the common origin in the centre of the star \citep[e.g.][Fig.~3.4a]{Lorimer2005}. 
This transformation does not change the subsequent derivations in the text. }
 \label{fig:geom}
\end{figure}

Two essential components of magnetospheric geometry are the inclination angle, $\alpha,$ and the impact angle, 
$\beta$. In general, $\alpha$ is defined as the angle between the spin and magnetic axes, and $\beta$ is the angle 
between the magnetic axis and the observer's LOS at the moment of closest approach. In this moment, the LOS passes 
through the plane defined by the spin and magnetic axes, called the fiducial plane (Fig.~\ref{fig:geom}). 

Unfortunately, there is no uniform convention for the range of possible values of $\alpha$ and the sign of 
$\beta$ in the published works \citep[see discussion in][hereafter EW01]{Everett2001}. For example, DR01 measured $\alpha$  from the rotational pole that is closer to the observer and is confined to 
the first quadrant: $0\degree<\alpha<90\degree$. For any $\alpha$, the impact angle, $\beta,$ is negative when the LOS 
passes ``polewards'' from the magnetic axis (``inside traverse'') and positive otherwise (``outside traverse'').

A more uniform convention was adopted
by EW01; this is used in this paper as well. The geometry in EW01 
is determined by the inclination angle, $\alpha$, and the viewing angle, $\zeta$, which is the angle between the 
spin axis and the LOS. Both $\alpha$ and $\zeta$ are measured clockwise from the direction of the angular 
momentum vector of the pulsar and can have any value between $0\degree$ and $180\degree$. The impact angle, $\beta,$ is measured 
clockwise from the direction of magnetic axis and is therefore positive when $\zeta>\alpha$ and negative otherwise.

\subsection{Derivation of geometry in DR01}    

\citet{Deshpande2000} started their derivation by analysing the frequency dependence of the observed profile width $w(\nu)$, which 
can be related to the frequency-dependent opening angle of the emission cone $\rho(\nu)$ by the 
expression \citep{Gil1984}\begin{equation}
 \cos \rho = \cos \alpha \cos(\alpha+\beta) + \sin \alpha \sin(\alpha+\beta)\cos(w/2).
 \label{eq:rho_w}
\end{equation}
The authors take the functional form of the $\rho(\nu)$ dependence from the study of \citet[][hereafter MD99]{Mitra1999},
who analysed the multi-frequency width measurements of 37 pulsars with both cone and core components. For these pulsars,
$\alpha$ was taken from the limiting width of the core 
component, $w=2.45\degree/(\sqrt{\pone}\sin\alpha)$ and $\beta$ from the maximum value of the PA gradient 
\citep{Rankin1990, Rankin1993b,Lyne1988}.  It is worth mentioning that for some of the pulsars in the sample of MD99, different $\alpha$ and 
$\beta$ were later obtained by EW01. 
MD99 fitted their multi-frequency $\rho(\nu)$ measurements with the relation:
\begin{equation}
 \rho(\nu)=\rho_\infty(1+K\nu_\mathrm{GHz}^{-a})\pone^{-0.5}.
 \label{eq:rho_nu}
\end{equation}
Here $\rho_\infty$ is the opening angle of the emission cone at infinite radio frequency, and $\pone^{-0.5}$ is the 
inverse square root of the pulsar period \citep[see also][]{Rankin1993a}. 
\begin{figure}
   \centering
\includegraphics[scale=0.75]{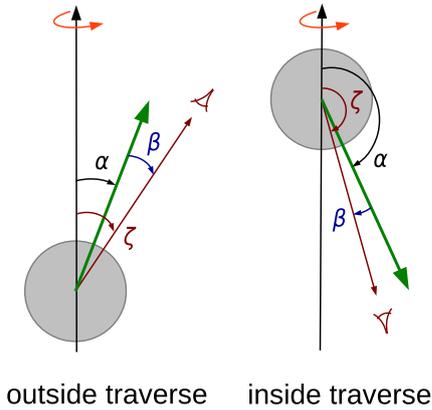}
 \caption{Pulsar orientation angles for the two branches of our 
 geometry fit. The pulsar rotates counter-clockwise, and the spin axis is shown by the vertical arrow. The shorter green 
 arrows show the magnetic axis, and the brown arrows show the LOS. The branches are labelled ``outside traverse'' and
 ``inside traverse'' to match the notation of DR01; ``inside''/``outside'' refers to the relative position 
 of the LOS direction  with respect to the smaller angle between the spin and magnetic axes. We follow the definition 
 of ($\alpha$, $\beta$) in EW01, which is different from the one used by DR01. The conversion between the two notations 
 is given in Table~\ref{table:angles}. }
 \label{fig:traverse}
\end{figure}

\begin{table}
\caption{Relationship between the conventions for inclination angle $\alpha$ and impact angle $\beta$
in DR01 and EW01 for the ``outside traverse'' and ``inside traverse'' branches of the geometry fit in DR01.}             
\label{table:angles}      
\centering                        
\begin{tabular}{|c | c | c |  }     
\hline
\multicolumn{3}{|c|}{\rule{0pt}{2ex} outside traverse} \\
\hline
\rule{0pt}{2ex}  $\alpha_\mathrm{DR}>0\degree$ & $\alpha_\mathrm{EW}<90\degree$ & $\alpha_\mathrm{DR}=\alpha_\mathrm{EW}$ \\
                  $\beta_\mathrm{DR}>0\degree$  & $\beta_\mathrm{EW}>0\degree$ & $\beta_\mathrm{DR}=\beta_\mathrm{EW}$  \\
                 \hline
\multicolumn{3}{|c|}{\rule{0pt}{2ex} inside traverse} \\
\hline
\rule{0pt}{2ex} $\alpha_\mathrm{DR}>0\degree$  & $\alpha_\mathrm{EW}>90\degree$ &  $\alpha_\mathrm{DR}=180\degree - \alpha_\mathrm{EW}$  \\
                  $\beta_\mathrm{DR}<0\degree$  & $\beta_\mathrm{EW}>0\degree$   &  $\beta_\mathrm{DR}=-\beta_\mathrm{EW}$  \\             
\hline
\end{tabular}
\end{table}

\citet{Deshpande2000} defined $w$ as the distance between the outer half-maximum points on the average profile. The authors performed
least-squares fits of Eqs.~\ref{eq:rho_w} and \ref{eq:rho_nu} to seven measurements of $w$ at frequencies between 
25 and 111\,MHz. They used a set of trial $\alpha$, $\beta$ and $a$, while leaving $K$ fixed at the best-fit 
value from MD99, $K=0.066$. The opening angle of the emission cone at infinite frequency was set by 
$\rhoG=\rho_\infty\times1.066\times\pone^{-0.5}$ at the values for the inner and outer cone from \citet{Rankin1993a}, 
namely at $\rhoG=4.3\degree$ and $5.7\degree$, respectively (B14 erroneously quoted these values as coming 
from MD99). 

The fit resulted in three solution branches: ``inside traverse'', ``outside traverse'', and ``pole-in-
cone'' models. For all three models the choice of $\rhoG$ had little influence on the quality of the fit. Quoting 
DR01, ``it appears that models with any reasonable 1-GHz $\rho$ values will behave similarly''.

The best-fit $\alpha$ and $\beta$ from all three models can then be used to calculate the swing 
of the linearly polarised radiation position angle
 under the assumptions of the rotating vector model \citep[RVM,][]{Radhakrishnan1969},
and the results can be compared to the observations. For this comparison, DR01 used the absolute 
value of the maximum gradient of the PA angle 
$|\RPA|=|\sin\alpha/\sin\beta|$ (see also \citet{Rankin1993a} and the discussion in EW01). 
The inside and outside traverse models resulted in $|\RPA|$ similar to the observed 
value, whereas the pole-in-cone model yielded an incompatible $\RPA$ and thus was rejected.

The inside traverse model gave the following pairs of geometry angles: $\alpha_\mathrm{DR} = 15.39\degree$ and 
$\beta_\mathrm{DR} = -5.69\degree$ for the outer-cone $\rhoG$, and $\alpha_\mathrm{DR} = 11.58\degree$, 
$\beta_\mathrm{DR} = -4.29\degree$ for the inner-cone $\rhoG$. The values for the outside traverse model were similar, 
but with positive $\beta$.  The outside and inside traverse geometries prescribed different behaviour of the 
magnetic azimuth $\psi(\phi)$, since for the small $\phi-\phi_0$, it is proportional to $\sin(\alpha+\beta$) (see 
Sect. ~\ref{sec:phtr}). The rate of magnetic azimuth change can be externally constrained if the number of sparks 
in the carousel $N$ is known: the change in magnetic azimuth between the adjacent subpulses detected within the same 
pulse period should be equal to $360\degree/N$. DR01 deduced $N=20$ from the $37\pone$ modulation 
of the drifting subpulse amplitudes, observed in a subset of their pulse stacks (see Sect.~\ref{sec:no20}). 
The inside traverse model predicts 
the correct number of sparks, whereas the outside traverse model predicts approximately eight sparks and therefore is 
rejected.

\begin{figure*}
   \centering
 \includegraphics[scale=0.75]{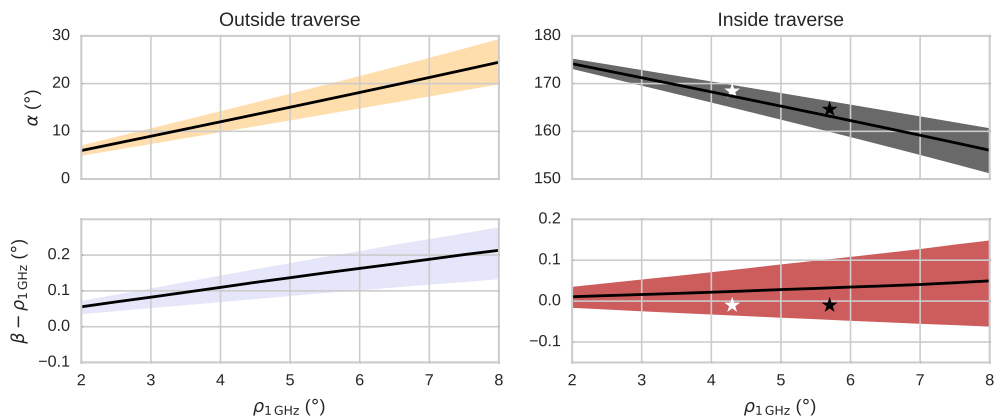}
 \caption{Pulsar orientation angles based on the analysis of frequency-dependent component separation and the 
 PA sweep rate versus the assumed value of the emission cone opening angle at 1\,GHz, $\rhoG$. The black lines 
 show $\alpha(\rhoG)$ and $\beta(\rhoG)$ for $\RPA=-2.9$\,$\degree/\degree$. Shaded regions mark a $\pm0.5$\,$\degree/\degree$ 
 uncertainty in $\RPA$, where the lower edge corresponds to a lower absolute value of $\RPA$. The stars mark the 
 DR01 geometry solutions corresponding to the inner and outer cone $\rhoG$ (white and black, respectively). 
 The angles were converted into the EW01 notation according to Table~\ref{table:angles}.  }
 \label{fig:alphabeta}
    \end{figure*}

\begin{figure}
   \centering
 \includegraphics[scale=1.1]{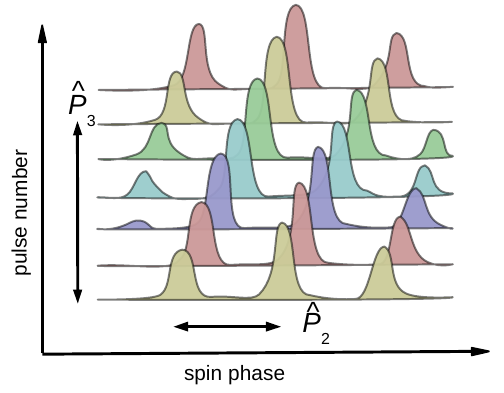}
 \caption{Example of a drifting subpulse sequence. $\hptwo$ corresponds to the observed longitudinal spacing between 
 subpulses, and $\hpthr$ is the amount of time between subpulses that appear at the same longitude. Here, $\hpthr$ is 
 equal to the integer number of the rotation periods, but this is not necessarily true in general.  }
 \label{fig:P2P3}
\end{figure}

\subsection{Geometry fit with LOFAR observations}

We repeated this analysis based on 45 $\phi_\mathrm{sep}$ values from four observing sessions in B14. These 
measurements were obtained by fitting two Gaussian components to the session-integrated average B-mode profiles at frequencies between 26 and 80\,MHz. We chose to fit the separation between components, not the full width at half-maximum
(FWHM), since the latter is affected by smearing at the low edge of the band (below 40\,MHz) due to the incoherent de-dispersion (B14).
At the same time, using $\phi_\mathrm{sep}$ facilitates comparison with MD99, who performed their analysis for the radius 
of the maximum intensity inside the cone. 

Similarly to DR01, we find two solution branches: one with $\alpha<90\degree$ , and another with $\alpha>90\degree$, 
corresponding to outside and inside traverses in their notation (see Fig.~\ref{fig:traverse}). Since we did not detect 
a $37\pone$ modulation of the subpulse amplitudes (Sect.~\ref{sec:no20}), we chose not to reject the outside traverse branch.

The outside and inside traverse models constrain $\beta$ quite well, but give comparable values of $\chi^2$ for a wide 
range of $\alpha$. To constrain $\alpha$ further, we used $\RPA=-2.9\pm0.5$\,$\degree/\degree$  from
\citet{Suleymanova1998}. This range reflects the likely uncertainty on $\RPA$: other works also give 
$\RPA=-2.7$\,$\degree/\degree$ (DR01) and $-3.0$\,$\degree/\degree$ \citep{Backus2010}. Similarly to EW01, we used the 
following expression for the $\RPA$:
\begin{equation}
 \RPA\equiv\left.\dfrac{d\mathrm{PA}}{d\phi}\right\vert_{\phi=\phi_0} = -\dfrac{\sin\alpha}{\sin\beta},
 \label{eq:PA}
\end{equation}
which fixes a widespread error regarding the direction of position angle growth on the sky: the classical 
expression for the PA swing in RVM was derived assuming that the PA increases clockwise on the sky, whereas the actually 
measured PA increases counter-clockwise on the sky \citep[following the standard astronomical convention, see for 
example][]{vanStraten2010}.

\begin{figure*}
   \centering
 \includegraphics[scale=0.75]{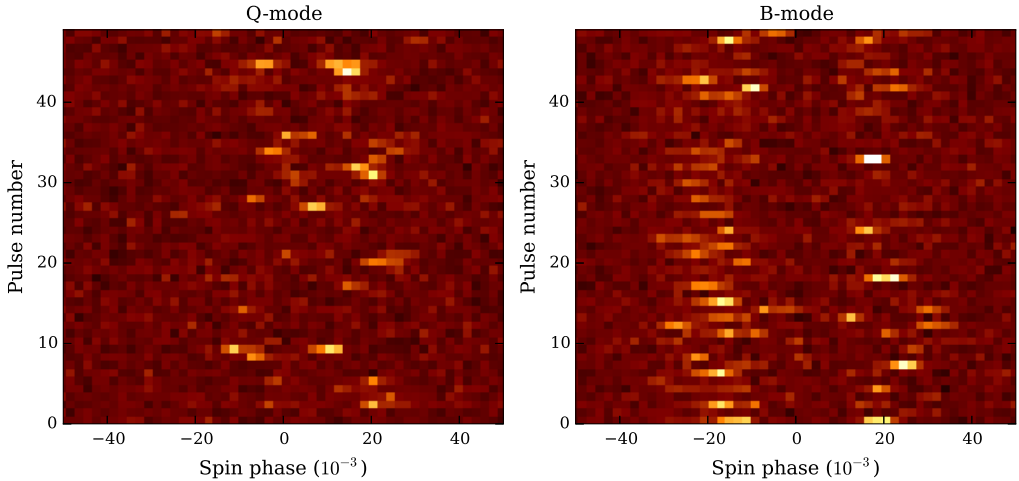}
 \caption{Example of pulse sequences in the Q and B modes from the observing session L102418. Only the on-pulse longitude 
 region is shown. 
 In the B mode the apparent  drift period $|\hpthr|$ is similar to two rotation periods, which results in the distinct even-odd 
 modulation of the subpulses (see Sect.~\ref{sec:Bsubp} for the discussion of the $\hpthr$ sign). If  $|\hpthr|$
 were exactly 2$\pone$, drifting subpulses would appear at the same longitudes during every other rotation period.
 Because $|\hpthr|$ is slightly larger than $2\pone$, however,
the subpulses appear to be slowly shifting towards the leading edge of
 the average profile, with an  apparent cycle of about $10\pone$. }
 \label{fig:ps}
\end{figure*}

In contrast do DR01, the outside traverse resulted in a better fit in our case, with a reduced $\chi^2$ of about 2.7, while the 
inside traverse gave about 3.9. The fit was consistently better at steeper values of $\RPA$, but we did not try 
to minimise the $\chi^2$ by varying $\RPA$. We note that fitting for $K$ in Eq.~\ref{eq:rho_nu} and with better
measurements of PA may result in better $\chi^2$, but exploring this is beyond the scope of the current analysis. 
It is also possible that the errors on $\phi_\mathrm{sep}$, set by the two-Gaussian fit in B14, are underestimated.

Similarly to DR01, we found that the quality of fit was the same for the inner and outer cone values of $\rhoG$.
Considering that $\rhoG$ has a direct influence on the derived ($\alpha$, $\beta$) and given the absence of any 
additional information about the opening angle of the \src\  emission beam \citep[although the steep spectral evolution 
of the separation between profile components may give preference to the outer cone, with $\rhoG$ close to $5.7\degree$,][]{Rankin1993a}, 
it seems reasonable to explore the possible 
range of ($\alpha$, $\beta$) values under the assumption that $\rhoG$ lies somewhere within the values found in MD99. 
Thus, we repeated the fit for a set of trial $2\degree<\rhoG<8\degree$, loosely corresponding to the 
$2.9\degree<\rhoG<7.3\degree$ span of calculated $\rhoG$ in the sample of pulsars in MD99 for $\pone\approx1.1$\,s.
The ($\alpha$, $\beta$) for different $\rhoG$ are plotted in Fig.~\ref{fig:alphabeta}. The errors on ($\alpha$, 
$\beta$) are set by the uncertainties in $\RPA$. The power-law index in Eq.~\ref{eq:rho_nu}, $a$,
was 0.45 for outside traverse and 0.33 for inside traverse. 
As a rule of thumb, for any chosen $\rhoG$, $\beta \approx \rhoG$,  in line with DR01's statement that the LOS
grazes the edge of the emission cone at frequencies $\gtrsim1$\,GHz. The inclination angle $\alpha$ for the outside 
traverse, and $180\degree-\alpha$ for the inside traverse, increases with $\rhoG$ approximately as $3\rhoG$.

\citet{Deshpande2000} give the numerical values of the geometry angles only for the inside traverse solution branch. For the inner 
and outer cone $\rhoG$, their ($\alpha$, $\beta$) are within our range of possible values (Fig.~\ref{fig:alphabeta}).

\section{Single-pulse analysis: data processing and drift parameterisation}
\label{sec:sps}
For the single-pulse analysis, the signal was de-dispersed, integrated in frequency within several sub-bands, and folded 
synchronously with the topocentric pulsar period to form frequency-resolved ``pulse stacks'', $s(\phi,t)$, 2D 
arrays of the uncalibrated total intensity as a function of rotational longitude and time. In this notation, time is assumed to be 
constant within each rotational period: $t=[$pulse number$]\times\pone$. 

The signal within each pulse period and sub-band was normalised by subtracting the mean and dividing by the standard 
deviation of noise in the off-pulse window. Pulse stacks from the B and Q modes were analysed separately, for which the 
few-second region around the mode transition was removed. The mode transitions were identified visually through abrupt 
(within a few rotation periods) start or cessation of the subpulse drifting patterns.

In this paper we adopt the somewhat simplified version of the drift parameterisation from the work of 
\citet[][hereafter ES03]{Edwards2003}. The authors
treated pulse stacks as the sum of steady and drifting components (the steady component 
can also depend on $\phi$ and $t$, but in aperiodic fashion). The drifting component can be expressed as 
\begin{equation}
 s_\mathrm{drift}(\phi, t) \sim A(\phi)\times \mathrm{Re}\left[\exp \left(2\pi it / \hpthr + i\theta(\phi)\right)\right].
 \label{eq:sdrift}
\end{equation}
Here, $A(\phi)$ is the variation in the pulse amplitudes within the on-pulse window, on average described by the shape of the average 
profile, $\hpthr$ is the observed modulation period along the lines of constant $\phi$, and $\theta(\phi)$ is the so-called 
``drift phase''. The drift phase accrues $360\degree$ between two adjacent subpulses that are recorded within the same pulse period. 
The observed longitudinal spacing between such subpulses is called $\hptwo$:
\begin{equation}
 \hptwo = \pone\left[\frac{d\theta}{d\phi}\right]^{-1}.
 \label{eq:HP2}
\end{equation}
A cartoon with drifting subpulses together with their respective $\hptwo$ and $\hpthr$ is shown in Fig.~\ref{fig:P2P3}. 

As noted by ES03, for the parameterisation given by Eq.~\ref{eq:sdrift}, the apparent direction of drift (i.e. whether pulses 
seem to march towards the leading or trailing edges of the profile) depends not on the sign of $\hpthr$ or 
$d\theta/d\phi$, but on whether they both have the same sign \citep[in general, the sign of $d\theta/d\phi$ is constant throughout 
the on-pulse window, although a few exceptions are known, see][]{Champion2005,Weltevrede2006,Szary2017}. 
If the sign of $d\theta/d\phi$ (and thus $\hptwo$) matches the sign of $\hpthr$, subpulses will appear earlier with each 
successive period, thus, drifting towards the leading edge of the average profile. In the case of opposite signs, subpulses 
will appear to be drifting to the trailing edge of the average profile. To avoid confusion, ES03 assumed that $d\theta/d\phi$
is always positive and let the sense of drift be determined by the sign of $\hpthr$.

The properties of subpulse drift are traditionally assessed through so-called longitude-resolved fluctuation spectra \citep[LRF 
spectra; see][DR01]{Backer1970a}. An LRF spectrum consists of 1D Fourier transforms of the pulse stacks, computed over the 
lines of constant rotational longitude. If the time-dependent subpulse modulation is present at the spin phase 
$\phi$, the Fourier transform 
of $s(t)$ for that $\phi$ will have two peaks at Fourier frequencies of $\nu_t=\pm 1/\hat{P}_3$. Since the peaks are present 
in both the positive and the negative half of the Fourier spectrum, it is not possible to determine the sign of $\hpthr$ from the power 
spectrum alone. However, the direction of the drift can be retrieved from the complex-valued Fourier spectrum by examining the 
evolution of the complex phase of the transform with the longitude. If $S(\phi, \nu_t)$ is the longitude-resolved Fourier 
transform of $s(t)$, then from Eq.~\ref{eq:HP2} it follows that
\begin{equation}
\begin{aligned}
\mathrm{Arg}\left[ S(\phi,\nu_t =1/\hpthr)\right] & \sim \thph, \\
\mathrm{Arg}\left[ S(\phi,\nu_t =-1/\hpthr)\right] & \sim -\thph. 
\end{aligned} 
\label{eq:LRF_peaks}
\end{equation}
Since $d \mathrm{Arg}[S]/d\phi>0$ by definition, selecting the peak with the observed positive phase gradient will fix the sign 
of $\hpthr$ (ES03).

\begin{figure*}[th]
 \centering
 \includegraphics[scale=0.95]{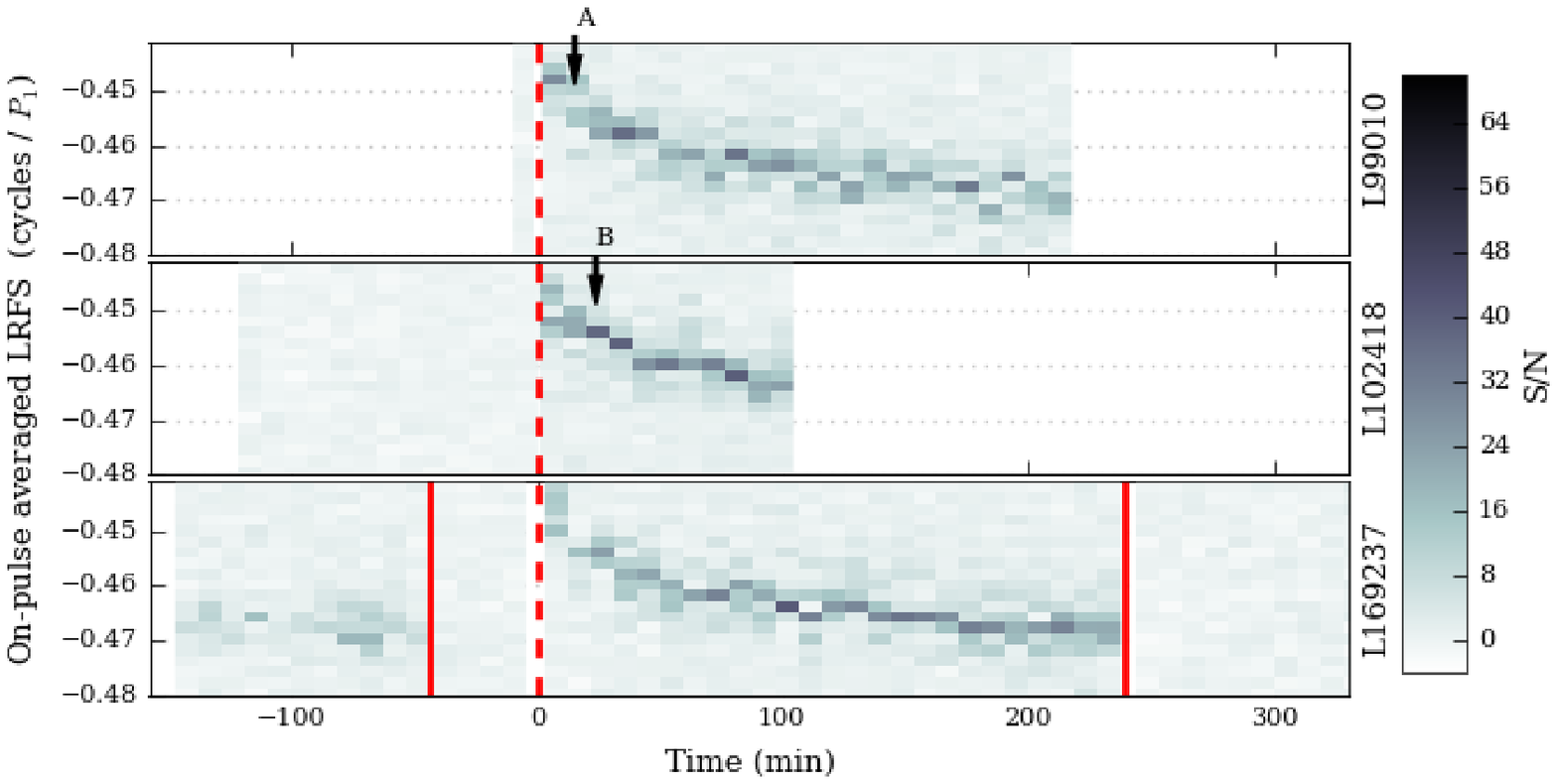} 
 \includegraphics[scale=0.95]{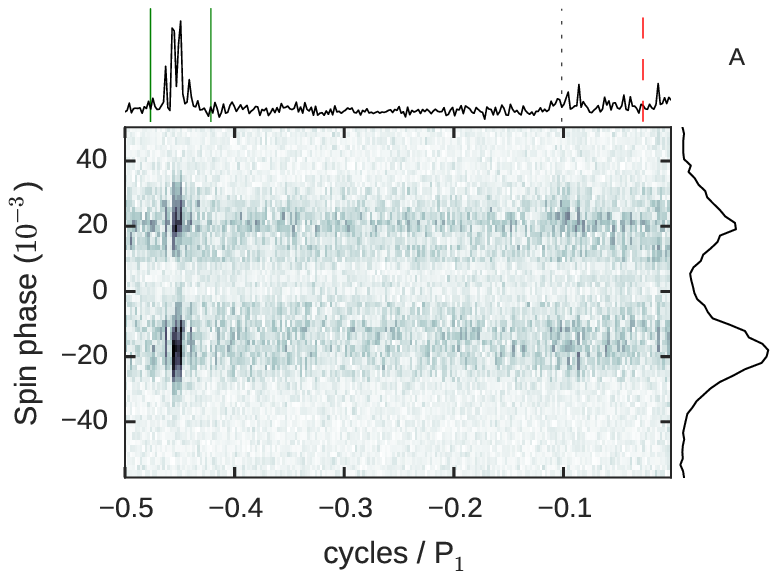}\includegraphics[scale=0.95]{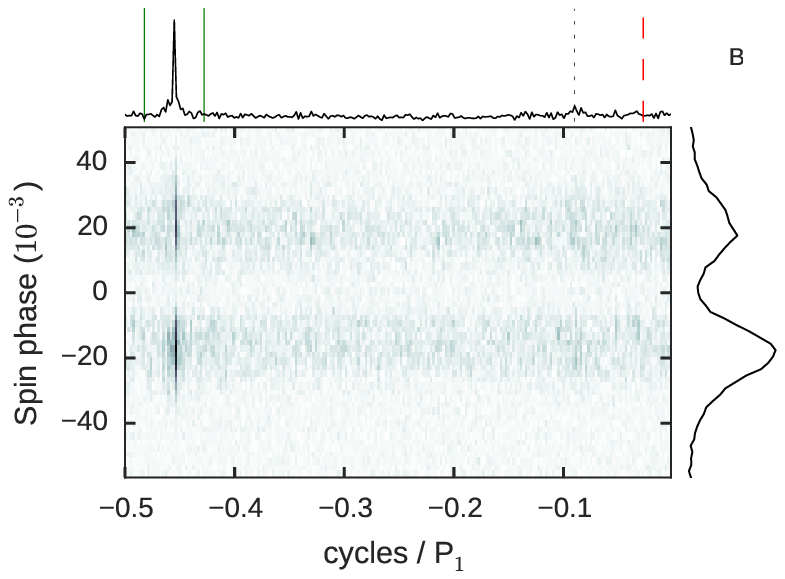}
 \caption{ \textit{Top plot:} Stack of LRF power spectra, integrated within the on-pulse window. On the X-axis, each column
 corresponds to the LRF spectrum computed from 512-pulse ($\approx 9$\,min) pulse stacks. Only the Fourier frequencies 
 around the B-mode feature are shown. The S/N ratio is shown with colour. Similarly to B14, observations are aligned by 
 the Q-to-B transition. Vertical solid lines mark B-to-Q transitions, and dashed lines show the Q-to-B transitions. 
 The observed $\hpthr$ slowly evolves with time since the B-mode onset and also exhibits strong random fluctuation 
 around the general trend.  \textit{Bottom row}: Two examples of individual LRF spectra showing different types of $P_3$ 
 behaviour: multiple peaks (A) and a single peak (B). The average profiles for the corresponding  pulse sequences are 
 plotted in the side panel. Two solid vertical lines mark the supposed position of the sidelobes (for the highest S/N peak in A), calculated 
 for the 20-spark carousel from DR01. The dotted line shows the first-degree alias of the second harmonic frequency of 
 the highest S/N peak, and the  dashed line marks the frequency of the supposed 37$\pone$ modulation from \citet{Asgekar2001}.}
 \label{fig:P3_vs_t}
\end{figure*}

\section{Searching for subpulse modulation in Q mode}
\label{sec:Qsubp}
Normally, LRF spectra of the Q-mode pulse stacks do not exhibit any prominent peaks,  
indicating that the subpulses arrive predominantly chaotically within the on-pulse 
longitude window \citep[Fig.~\ref{fig:ps}, left; see also DR01,][]{Rankin2003,Rankin2006b}. 
However, possible weak, intermittent, or narrowband periodicities have not been 
ruled out completely. \citet{Deshpande2000}, having examined a 170-pulse sequence right after a B-to-Q 
transition at 430\,MHz, found a weak (from a visual inspection of their Fig.~13, 
the LRF spectra averaged within the on-pulse window had a signal-to-noise
ratio (S/N) of about five), but distinct feature at the frequency of 0.77\,\cp. 
This result was not confirmed by a larger sample of five observations at 40 and 103\,MHz 
\citep{Rankin2003}. \citet{Rankin2006b} reported that the LRF spectra from a 30-minute-long 
sequence of the Q-mode pulses right before the Q-to-B transition in one of their three 
327-MHz observations  had a feature at $\nu_t=0.0275\pm0.001$\,\cp, which is similar to 
the B-mode carousel circulation frequency.

Our observations contained 4.2 hours of Q-mode pulse sequences and five mode transitions, forming 
a good dataset to search for any potential subpulse drift in the Q mode. First, we searched 
for intermittent and possibly narrowband periodicities. We analysed pulse stacks 
from three central $\approx 15$-MHz sub-bands (central frequencies of 39, 54, and 70\,MHz).
In these sub-bands, the Q-mode signal was strong enough to form a discernible average 
profile over the course of one session. We divided the pulse stacks into 512-pulse 
blocks and computed LRF spectra on each block separately. In order to reduce potential 
smearing of the spectral features by the slow variation of the
pulse intensities (for example, at 50\,MHz, the pulsar signal scintillates with a characteristic timescale of 4\,min, 
or 240 pulsar periods, see Fig.~1 and Appendix~B of B14), we subtracted a running 32-period 
on-pulse mean from the pulse sequences. None of the obtained LRF spectra (integrated in longitude 
within the on-pulse window) had features with an $\mathrm{S/N}>5.2$. The features 
with an $\mathrm{S/N}>4$ appeared to be randomly scattered in Fourier frequency, and none 
of them  fell within the 0.026--0.029\,\cp\ interval, corresponding to the range of observed
carousel circulation frequencies for the B-mode pulse sequences.

In order to increase our sensitivity to broadband periodicities
that are stable in time, we summed 
the pulse stacks from different sub-bands, re-computed LRF spectra on 512-pulse blocks, 
and added the spectra, forming one average LRF spectrum per an observing session. 
None of the these spectra had features with an $\mathrm{S/N}>3.8$.
We therefore conclude that we were unable to find any periodicities in the available Q-mode 
observations.

\begin{figure*}
   \centering
 \includegraphics[scale=0.9]{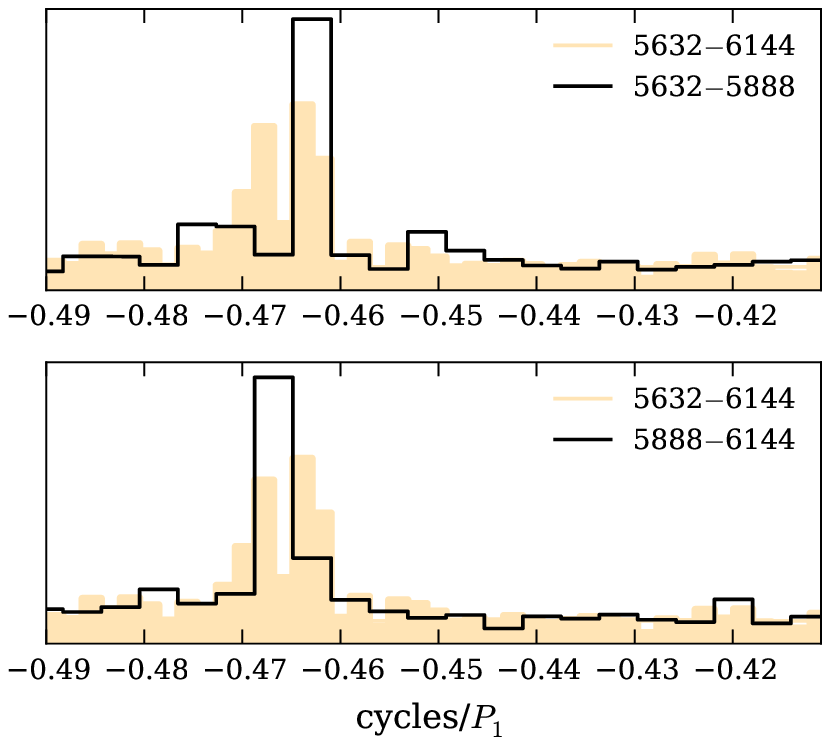}\includegraphics[scale=0.9]{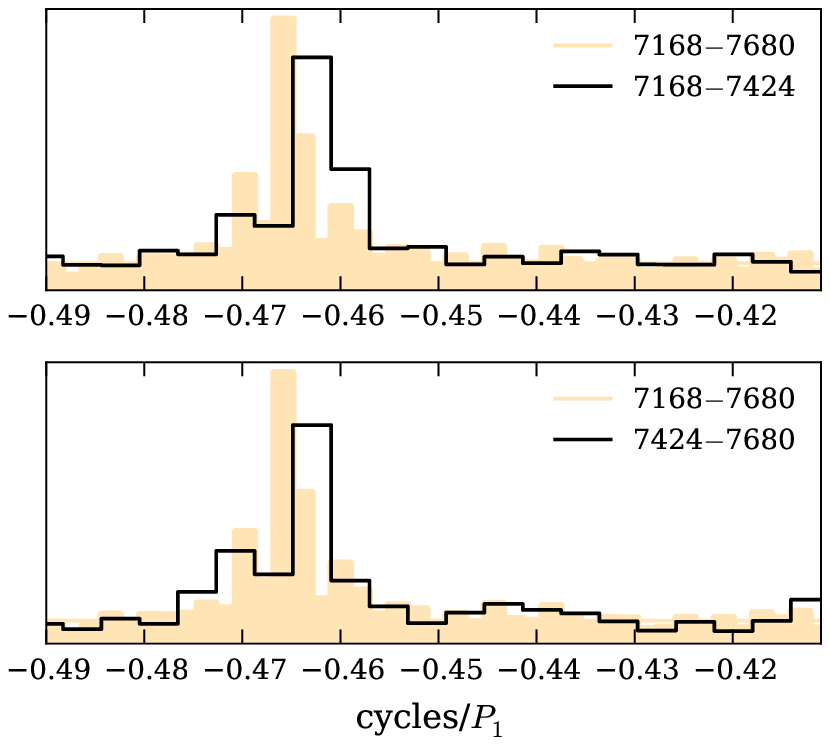}
 \caption{On-pulse averaged LRF spectra for two 512-pulse sequences (light colour, filled) 
 together with the LRF spectra computed for the first and last 256 pulses of each pulse sequence 
 (black, empty). Only the Fourier  frequencies around the B-mode feature are shown. The 
 pulse numbers in the legend are counted from the closest Q-to-B transition. \textit{Left:} 
 Example of an LRF spectrum with two peaks that are resolved into a single peak that changes its 
 position with time. All double features with clearly separated peaks were resolved in this way. 
 \textit{Right}: Example of an LRF spectrum for a 512-pulse sequence that shows the peak 
 offset from the peaks in the LRF spectra of the subsequences. This behaviour might be 
 explained by phase jumps within the pulse sequence.}
 \label{fig:compar}
\end{figure*}

\section{Subpulse modulation in B mode}
\label{sec:Bsubp}

\subsection{Overall properties}    

The remarkable drift of the \src\  subpulses (Fig.~\ref{fig:ps}, right) was noted soon after
the pulsar was discovered \citep{Taylor1971} and has been extensively studied since then 
\citep[see, e.g. the most recent work of][and references therein]{Backus2011}. 
Normally, the LRF spectra of B-mode pulse stacks present a strong feature at 
$\nu_t$\,$\approx$\,$\pm 0.46$\,\cp. The observed value of $|1/\hpthr|$ depends on the 
time since B-mode onset: right after Q-to-B transition, $|1/\hpthr| \approx 0.455$\,\cp\ 
and then it slowly relaxes to $\approx 0.47$\,\cp\ in approximately exponential fashion 
\citep{Rankin2006b,Backus2011}. $\hpthr$ is considered to be independent of observing 
frequency, although \citet{Rankin2003} noted small discrepancies in $1/\hpthr$ measured 
simultaneously at 40 and 103\,MHz. In their work, $1/\hat{P}_3$ refers to the 
interpolated centroid of the spectral feature, and the error comes from the formal uncertainty
of centroid estimation. The discrepancies were of the order of 0.001\,\cp\ or 
a  few assumed uncertainties. 

In order to investigate the subpulse periodicity in our B-mode observations, we computed 
LRF spectra on 512-pulse blocks, divided into five sub-bands in radio frequency. 
Only three central sub-bands (at 39, 54 and 70\,MHz) had a sufficiently
high S/N to discern 
a subpulse periodicity in each block. The LRF spectra had peaks close to $\approx \pm 0.46$\,\cp\ 
in the on-pulse window. Examination of $d\theta/d\phi$ from the positive and negative parts 
of the complex-valued Fourier spectrum (Eq.~\ref{eq:LRF_peaks}) revealed that the phase gradient 
is positive for the peak at $\nu_t<0$. Thus, $\hpthr<0$ and subpulses \emph{\textup{ostensibly}} 
drift to the trailing edge of the profile (see also Sect.~\ref{sec:fdphd}).
The true sense of drift may be different 
due to aliasing, see the discussion in Sect.~\ref{sec:phtr}.

Some of our LRF spectra exhibited simple narrow peaks, similar to those presented in 
DR01 or \citet{Rankin2003}. However, the majority of the LRF spectra appeared to have 
complex-shaped features with two or multiple peaks. Thus, we did not try to estimate 
the exact $1/\hpthr$ by fitting centroids to spectral features, as was done by \citet{Rankin2003}. 
Instead, we simply compared the positions of peaks between sub-bands. We found that 
for all 512-pulse blocks, the peaks in different subbands fell at the same $\nu_t$ bins, 
indicating that the instantaneous $1/\hpthr$ is stable across the LBA frequency range at 
least to within our spectral resolution of 0.004\,\cp. For the
subsequent analysis, we integrated 
the signal in the whole band to increase the S/N and re-computed the LRF spectra in the 
whole band. 

The upper panel of Fig.~\ref{fig:P3_vs_t} shows the near-peak portion of these on-pulse-integrated 
LRF spectra for all of the 512-pulse blocks for the three observing sessions. This panel 
also shows a part of the Q-mode LRF spectra (integrated within the respective Q-mode on-pulse window).
To highlight the secular evolution of $\hpthr$, the sessions were aligned by  Q-to-B transitions.
Our observations confirm the previously established secular evolution of $1/\hpthr$ \citep{Rankin2003,Rankin2006b,Backus2011}.
However, it is obvious that the spectra exhibit a variety of structures. Figure~\ref{fig:P3_vs_t} 
(lower panels) shows the $\nu_t<0$ region for a pair of LRF spectra featuring two qualitative 
shapes of LRF features: multiple peaks (A), and a single peak (B). Previously published works acknowledged this behaviour, 
although no special attention has been paid to it. \citet{Suleymanova2009} 
mention two discrete jumps in the feature by 0.007\,\cp\ within a 700-pulse long sequence of pulses 
right after a Q-to-B transition. \citet{Rosen2008} report additional  peaks of lower S/N on the 
sequence from \citet{Suleymanova1998}, separated by about 0.003\,\cp. \citet{Rankin2006b} also report 
a few double peaks in the LRF spectra, separated by 0.007\,\cp. In panel A in Fig.~\ref{fig:P3_vs_t}, 
the characteristic separation between peaks is about 4 bins, or 0.0078 \cp. This agrees with the
spread of $1/\pthr$ points in Fig.~3 of \citet[][also computed  based on 512-pulse sequences]{Backus2011} 
and is equivalent to the spread of carousel circulation times in \citet{Rankin2003}, obtained 
from 256-pulse LRF spectra. The multiple peaks in LRF spectra are present throughout all our B modes. 

As was pointed out by \citet{Rosen2008}, observing two values of $\hpthr$ \emph{\textup{simultaneously}} would 
serve as a strong corroboration to the non-radial surface oscillation model of drifting subpulses. Unfortunately,
our ability to resolve the true temporal simultaneity is limited by the frequency resolution of the 
discrete Fourier transform, which is inversely proportional to the length of the pulse sequence. 
Nevertheless, we recomputed LRF spectra using the 256-pulse sequences and compared each pair with the 
corresponding 512-pulse LRF spectrum. For all 512-pulse LRF spectra with two distinct peaks, only 
one peak at a time was found in the 256-pulse subsets (see Fig.~\ref{fig:compar}, left), indicating 
that $\hpthr$ can change on a timescale as short as $256\pone$. 

Multiple peaks in LRF spectra may partly be attributed to phase jumps, caused by momentary shifts 
in spark positions. As noted by \citet{Halpern2017}, a sudden large (of the order of $\pi$) jump 
in phase of a single-frequency periodic signal can result in an apparent frequency splitting of 
its power spectrum, with the two new peaks straddling the frequency of the absent primary peak. 
We have searched for such frequency splitting in our observations by comparing the LRF spectra from 512- 
and 256-pulse sequences. In some cases we found that the pairs of 256-pulse LRF spectra have distinct 
peaks at the same $\nu_t$, which did not coincide with the peak of the 512-pulse LRF spectrum 
(Fig.~\ref{fig:compar}, right), although we did not detect a symmetric splitting corresponding to a 
single phase jump. A simple simulation of a sinusoidal signal with a number of various phase jumps 
showed that adding more jumps to the signal can
result in uneven amplitudes of the split peaks, with one of the peaks absorbing most of the power. 
 
\begin{figure}
   \centering
 \includegraphics[scale=1.0]{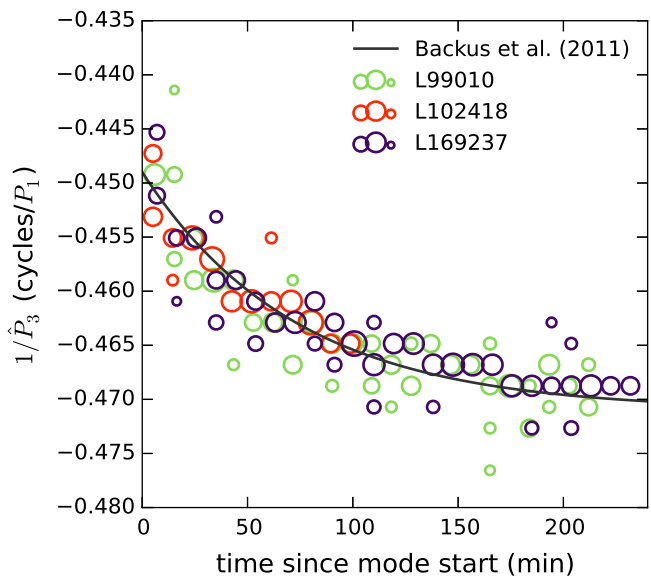}
 \caption{Primary feature of the LRF spectra from Fig.~\ref{fig:P3_vs_t} for the three B-mode intervals
 with observed Q-to-B transitions. If the LRF spectrum had several peaks, all of them
 were plotted. The size of the marker corresponds to the peak S/N. The fit of Eq.~\ref{eq:F3B11}
 from \citet{Backus2011} is overplotted, which was based on two partial B modes observed at 325\,MHz. 
  }
 \label{fig:F3Backus}
    \end{figure}

Although the true drift frequency may vary on short timescales or be obscured by phase jumps, there 
is little doubt in its gradual evolution throughout the B mode \citep{Rankin2006b}. We compared the mode-long 
evolution of $\pone/\hpthr$ from our observations to the most recent similar study of \citet{Backus2011}, who 
observed two partial B modes with the GMRT at 325\,MHz. They obtained the following 
expression for the $\hpthr$ evolution:
\begin{equation}
-\pone/\hpthr = 0.471-0.022\times \exp(-t/73),
\label{eq:F3B11}
\end{equation}
where $t$ is the time in minutes from the onset of the B mode. As noted 
by the authors, the functional form of the expression is completely arbitrary, and the main purpose of the fit 
was  to gauge the characteristic time of the parameter evolution in the B mode. Equation~\ref{eq:F3B11}
is consistent with our measurements as well (Fig.~\ref{fig:F3Backus}), indicating that 
the general evolution of $\hpthr$ with time from the mode start does not change noticeably 
between 40 and 325\,MHz and on a timescale of a few years, which separates our observations from those of 
\citet{Backus2011}.

\subsection{Search for periodic amplitude modulation in drifting subpulses}    
\label{sec:no20}

Periodic amplitude modulation of drifting subpulses, reported in the work of DR01, was proven to be 
a vital argument for constraining several important system parameters: the number of sparks in the 
carousel, the degree of aliasing, and the orientation of the LOS vector with respect to magnetic and spin 
axes. According to DR01, this tertiary modulation manifested itself in the form of symmetric sidelobes
 in the LRF spectrum, 0.0268\,\cp\ away from the primary feature at 0.4645\,\cp. These sidelobes, 
interpreted as an indicator of the persistent pattern in subpulse brightness, were detected only in a 
small, 256-pulse fraction of one 430-MHz B-mode observation. 

Following ES03, the period of the tertiary modulation, $P_4$, can be expressed through $\hpthr$, together with the
integer number of sparks, $N$, and the degree of aliasing, $n$,
in the following way:
\begin{equation}
 \dfrac{P_4}{\pone} = \dfrac{N}{\left| n + \pone/\hpthr \right|}.
 \label{eq:P4}
\end{equation}
In the work of DR01, $N=20$ and $n=-1$, satisfying Eq.~\ref{eq:P4} for
$1/P_4 = 0.0268$\,\cp\, and $1/\hpthr=0.4645$\,\cp. 

ES03 also noted that other 
solutions to Eq.~\ref{eq:P4} exist, with $|n|>1$, for example $n=2$ and $N=92$, or $n=-3$ and $N=132$. 
However, for the $\alpha$ and $\beta$ derived using the DR01 method (Sect.~\ref{sec:geom}), 
such solutions do not agree with the observed slope of the drift phase tracks, 
$d\theta/d\phi=34$\,$\degree/\degree$ (DR01, see also Sect.~\ref{sec:phtr}). The same reasoning can be 
used to reject a periodic spark brightness pattern, for example, sparks being brighter in the opposite 
quadrants of magnetic azimuth. In this case, $N$ obtained through Eq.~\ref{eq:P4} should have been 
multiplied by the number of the brightness pattern repetitions.

\begin{figure*}
\centering
\includegraphics[scale=1.0]{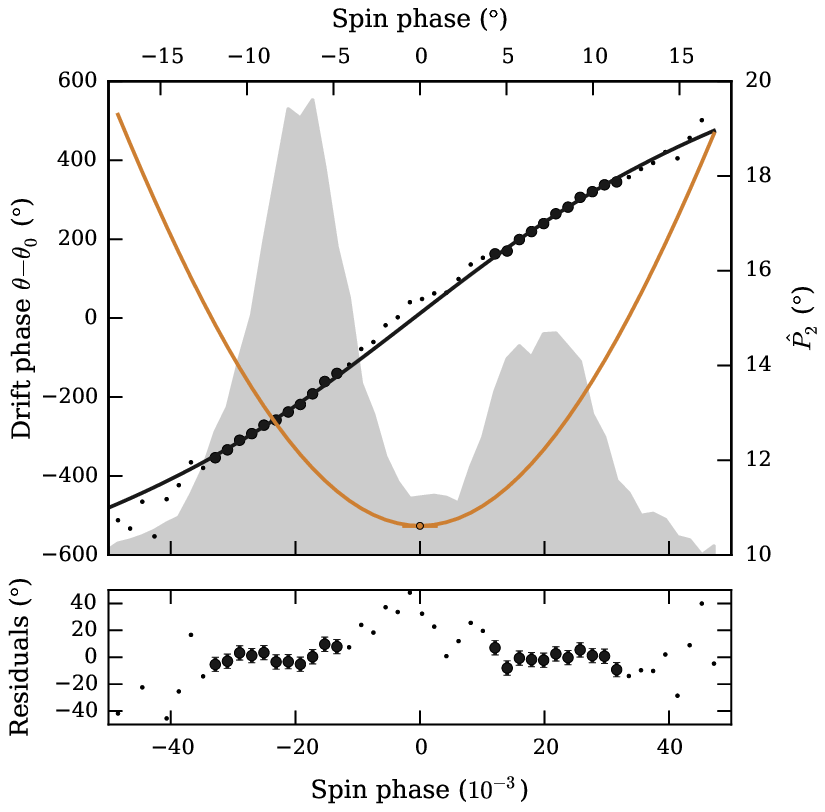}\includegraphics[scale=1.0]{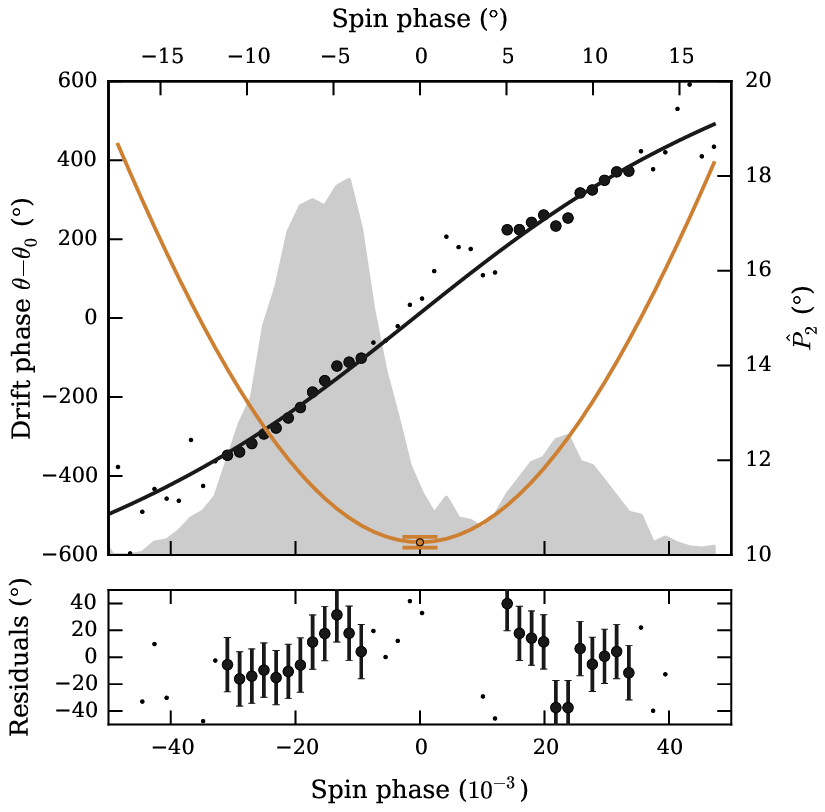}
\caption{ Unwrapped phase tracks, $\thph$ from Eq.~\ref{eq:LRF_peaks} (black dots) at the Fourier frequencies 
 corresponding to the $\pone/\hpthr$ peaks in two LRF spectra with relatively strong (left) and weak (right) 
 drifting features. The $\thph$ from the longitude region around the peaks of the average profile (larger markers) 
 were used to fit the number of sparks in the carousel according to Eq.~\ref{eq:theta}. Black lines mark an 
 example of such fit performed with the standard set of geometry/aliasing parameters (see text for details),
 and the bottom panels show the residuals together with the fitted  $\thph$ uncertainty. The orange line depicts 
 the longitudinal separation between subpulses, $\hptwo(\phi)$, computed  by differentiating the fit line 
 (Eq.~\ref{eq:HP2}).  The orange error bar at $\phi=0$ marks the $\hptwo$ uncertainty derived from the uncertainty 
 in fitted $N$. The average profile of the pulse sequence is shown with shades of grey. For convenience, 
 since $\hptwo$ is  measured in degrees, the top axis shows the spin longitude $\phi$ in the same units.}
 \label{fig:phtr}
\end{figure*}

\citet{Rosen2008} argued that the sidelobes may be due to the stochastic variations in the amplitudes 
alone. They analysed the same observation as in DR01 and found that sidelobes are present in the LRF 
spectrum, but are absent from the harmonic-resolved fluctuation (HRF) spectrum of the same 256-pulse 
sequence. An HRF spectrum,
a different manner of presenting a Fourier transform of drifting subpulses,  
is a stack of $1/\pone$ sections of the Fourier transform taken on the entire 1D pulse sequence
(see DR01). The authors performed Monte Carlo simulations of sidelobe significance by shuffling the 
pulse amplitudes in the sequence. The simulations showed that the chance probability of two peaks 
with the S/N seen in the LRF spectrum is quite low (lower than 1\%), but the probability of the modulation 
appearing in LRF and being masked by noise in HRF spectrum is equally low.

Further confirmation of sidelobes came from \citet{Asgekar2001} and \citet{Backus2010}. 
\citet{Asgekar2001} observed \src\ with the Gauribidanur Telescope for two sessions, each 
$\gtrsim1000$\,s long. The observations were conducted at 35\,MHz, with a 1 MHz bandwidth. In one of 
the sessions, the authors detected a peak at 0.027\,\cp\ (corresponding to the carousel circulation time 
in DR01), but without sidelobes around the primary feature. This low-frequency peak has about the same 
amplitude as the primary feature and appears in both LRF and HRF spectra. However, the spectrum is 
more noisy at frequencies below 0.1\,\cp\ , and another peak also lies at the even lower frequency of 
about 0.01\,\cp, which was not addressed by the authors. No low-frequency features were detected in the other 
session.

\citet{Backus2010} analysed 18 hours of Arecibo B-mode observations at frequencies of 327 and 
430\,MHz and found two more instances of tertiary modulation, again in the form of sidelobes around 
the primary feature. In one sequence, the sidelobes were quite weak, with a visually estimated $\mathrm{S/N}$ 
of about 3. The other sequence had more distinct sidelobes that
were asymmetric in amplitude. 
Unfortunately, the published LRF spectra do not allow inspecting
the $\phi$-resolved distribution
of power in the sidelobes. The authors state that modulation is rare, appearing in 5\% of the time span of the 
observations, and that they did not detect sidelobes corresponding to any other possible 
values of $P_4$. 

For the LOFAR observations, none of our LRF spectra (both in individual sub-bands and band-integrated) exhibited 
sidelobes corresponding to the 20-fold modulation of carousel sparks. We did not detect sidelobes for $N=20$, nor for any other $N$ between 5 and 40 (we note that this separation is much larger 
than the primary feature jitter in the previous section, see Fig.~\ref{fig:P3_vs_t}).  No obvious 
modulation features at low Fourier frequencies were found, although in some cases, we detected occasional 
peaks below 0.013\,\cp\, or the power was uniformly rising at lower frequencies in a red-noise-like manner.

Our dataset comprises 11.6 hours of B-mode observations, which
makes it 1.5 times smaller than the dataset of 
\citet{Backus2010}, which had two instances of tertiary modulation. Thus, it is possible that we did 
not detect the modulation solely because of its rarity. In addition, so far, the Gauribidanur and 
Arecibo detections of tertiary periodicity both came from relatively narrow-band observations that 
probed a small range of emission heights. In our observations the signal is collected from a greater 
range of heights, which means that if any persistent brightness pattern is frequency dependent, it might be 
smoothed out.

An independent confirmation of the tertiary amplitude modulation would be very important not 
only for constraining the system geometry and the parameters of the carousel, but also for distinguishing 
between the carousel and surface oscillation models of drifting subpulses \citep{Rosen2008}. The 
prospects of further investigations in this direction are quite good: recent observations of \src, 
conducted within the project of studying the simultaneous X-ray/radio mode-switching resulted in a large 
amount of radio data at low and high radio frequencies, surpassing all previous works (including 
this one) by a factor of several \citep{Mereghetti2016}. However, given the apparent absence 
of tertiary modulation in our observations and the concerns expressed in \citet{Rosen2008},
we decided for now to abandon the tertiary modulation constraint and treat both $n$ and $N$ as free parameters.

As was noted by the referee, subtracting
the running 32-period on-pulse mean may suppress possible amplitude modulation due to the $\approx37$-period 
carousel rotation. Thus, we repeated the analysis in Sect.~\ref{sec:Qsubp}--\ref{sec:Bsubp}
without subtracting the running mean, which yielded essentially the same results.
    
\section{Phase tracks in B mode}
\label{sec:phtr}

The complex argument $\thph$ in Eq.~\ref{eq:LRF_peaks} can provide additional information about the 
subpulse drift. An example of this $\thph$ dependence (often called the ``phase track'' in the literature)
is shown in Fig.~\ref{fig:phtr}. The $\thph$ accrues $360\degree$ over $\delta\phi=\hptwo$, the 
longitudinal separation between two adjacent subpulses in the same pulse period. For \src, $\hptwo$ is 
close to the width of the whole on-pulse window (see also DR01 and \citealt{Backus2011}), meaning that 
only one or two supbulses are detected per $\pone$. This can be directly confirmed by visual inspection 
of the B-mode pulse sequence in Fig.~\ref{fig:ps} (right). The flattening of the phase tracks far from 
the fiducial longitude  indicates that the longitudinal separation between subpulses is somewhat larger 
(by about 40\%) at the edges of the on-pulse window.  In the carousel model, this $\phi$-dependent subpulse 
separation is attributed to the curvature of the line traced by the LOS vector.

As was demonstrated by ES03, for the carousel consisting of $N$ uniformly spaced sparks, $\thph$
 is given by the formula
\begin{equation}
\label{eq:theta}
 \thph =  -N\psi\,\mathrm{sgn}\,\beta +(\phi-\phi_0)\left(n+\frac{P_1}{\hat{P}_3}\right) + \theta_0. 
\end{equation}
Here, the first term reflects the relationship between the change in magnetic azimuth ($\delta\psi=360\degree/N$)
and the corresponding change in drift phase ($\delta\theta=360\degree$) for the two adjacent sparks in a 
non-rotating carousel. Magnetic azimuth $\psi$ is expressed through rotational longitude  as
\begin{equation}
\label{eq:psi}
 \tan \psi = \frac{\sin(\phi-\phi_0)\sin\zeta}{\cos\zeta\sin\alpha-\cos(\phi-\phi_0)\sin\zeta\cos\alpha}.
\end{equation}
For the small $\phi$, $\alpha,$ and $\zeta$, this equation reduces to $\psi \approx\zeta\phi/(-\beta)$. 
In the notation of ES03, $d\theta/d\phi$ is always positive, thus a factor of $-\mathrm{sgn}\,\beta$ appears 
in the first term of Eq.~\ref{eq:theta}. The last two terms in Eq.~\ref{eq:theta} account for the carousel 
rotation around the magnetic axis and positions of sparks with respect to the fiducial longitude $\phi_0$.

At the fiducial longitude, $\hptwo$ is expressed as 
\begin{equation}
\label{eq:P2}
 \hptwo(\phi_0) = \pone\left[N\mathrm{sgn}\,\beta\,\frac{\sin\zeta}{\sin\beta}+\left(n+\frac{\pone}{\hpthr}\right)\right]^{-1}.
\end{equation}

\begin{figure}
\centering
\includegraphics[scale=1.0]{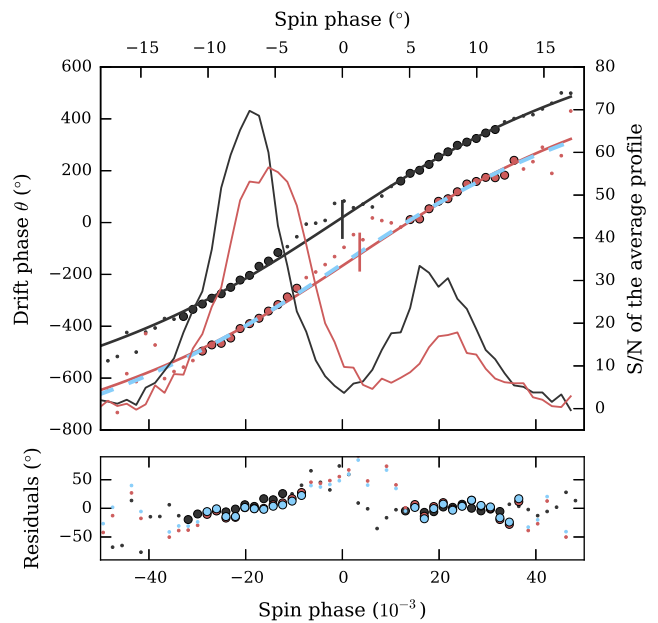}
\caption{Phase tracks for the beginning (black) and the end (red) of the same B mode (observation L99010), with an
artificial vertical offset added. The average profiles of the corresponding pulse sequences are plotted in the 
background with matching colours. The serifs mark the midpoint between profile components (see text 
for details). Lines show the fit with standard geometry/aliasing parameters. Solid lines correspond to 
the fit with fiducial longitude $\phi_0$ matching the midpoint between profile components. The light blue 
dashed line shows the fit to the late-mode phase track with $\phi_0$ fixed at the position of the midpoint 
between components at the onset of the B mode. Both constant and moving $\phi_0$ yield the same quality fits 
and an almost identical number of sparks $N$, and the change in $\phi_0$ is absorbed by fitted $\theta_0$. }
\label{fig:phtr_fid}
\end{figure}

Although Eq.~\ref{eq:theta} provides a specific mathematical prescription for $\thph$, using this equation 
directly for constraining $N$ and $\theta_0$ is difficult because of the number of unknowns it contains -- 
on the right-hand side,  only $\hpthr$ is measured directly. In the rest of this section we investigate how 
the best-fit $N$ depends on the chosen values for the other parameters in Eq.~\ref{eq:theta}, namely the 
degree of aliasing $n$, geometry angles, and the fiducial longitude $\phi_0$. We also explore the evolution 
of $N$ with radio frequency and the time since the B-mode start. For the latter we chose, with some degree of 
arbitrariness, a ``standard''  set of geometry/aliasing parameters: $\rhoG=4.5\degree$, 
$\RPA=-2.9\degree/\degree$, and outside traverse, $n=0$.

We limited the $\thph$ fit to the part of the on-pulse window around the average profile peaks:
$4\degree<|\phi-\phi_\mathrm{mid}|<12\degree$, where $\phi_\mathrm{mid}$ is the midpoint between the two Gaussians fitted to the profile 
components (B14). The fit was performed with a Markov chain Monte
Carlo (MCMC) 
algorithm\footnote{\url{https://github.com/pymc-devs/pymc}}. The measured $\thph$  was modelled as a normally 
distributed random variable, with the mean set by Eq.~\ref{eq:theta} and the fitted variance (same for all 
points within a single phase track) representing the unknown measurement error. All fitted parameters had uniform 
priors, and the quality of fit was verified by comparing model predictions to the observed $\thph$. The quoted 
errors on the fitted parameters represent 16th and 84th percentiles of their respective posterior distribution.

Although a carousel model in its simplest form prescribes an integer number of sparks, 
in our fit  we chose not to constrain $N$ to an integer number. As we show below, for any moderate degree of aliasing, 
a minor (with respect to their uncertainties) tweak in 
the geometry angles can always be used to bring $N$ to an integer. 
Also, \citet{Backus2011}
speculated about a possible flickering between configurations with a different number of integer sparks, which would 
produce seemingly non-integer $N$.

\begin{figure*}
 \centering
 \includegraphics[scale=0.95]{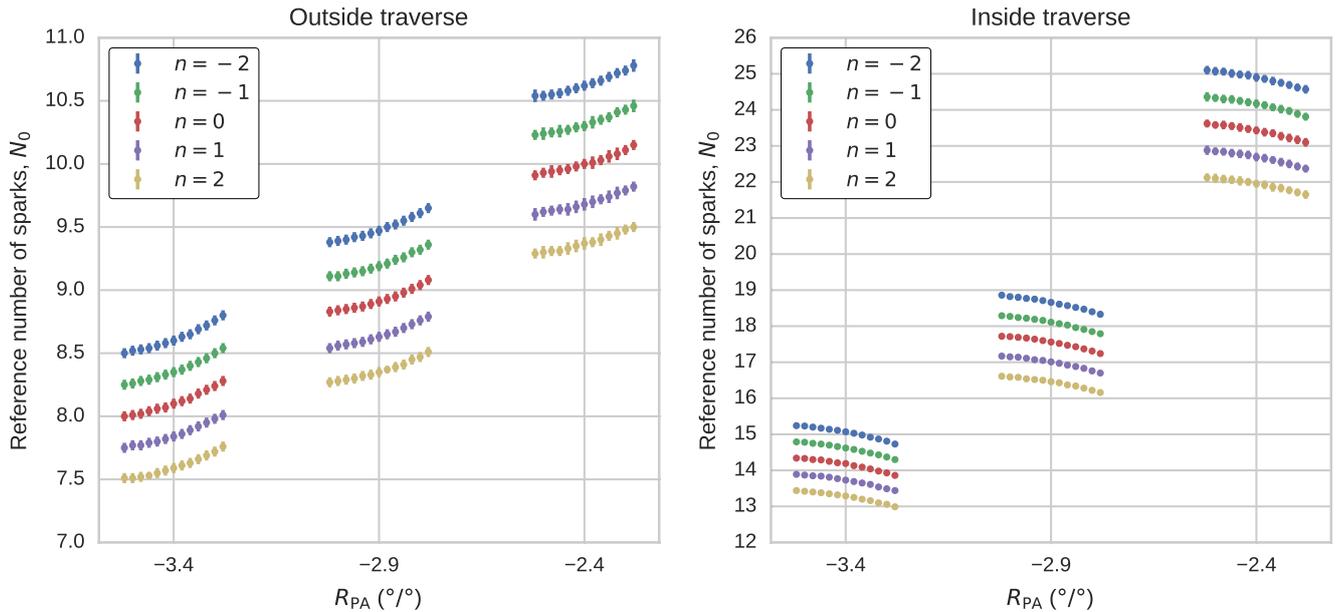}
 \caption{Reference number of sparks, $N_0$, under different assumptions of the geometry angles and the degree of aliasing, $n$. 
 $N_0$ was obtained by fitting Eq.~\ref{eq:theta} to a single pulse stack with high S/N in session L99010.
 The left and right panels show $N_0$ for the outside and inside traverse geometry branches. For each of the three 
 values of $\RPA$, namely $-3.4$, $-2.9$, and $-2.4$\,$\degree/\degree$, a set of trial $\rhoG$ between $2\degree$ and 
 $8\degree$ was used. For clarity, the points for different $\rho_0$ are plotted with the horizontal offset proportional 
 to $\rho_0-4.5\degree$.  Although the carousel model prescribes the integer number of sparks, $N_0$ was not fixed to an integer 
 value during the fit.
 }
 \label{fig:N_geom}
    \end{figure*}

\subsection{Dependence on the choice of fiducial longitude  $\phi_0$}
\label{subsec:phi0}

The rotational longitude  $\phi$ enters Eqs.~\ref{eq:theta} and \ref{eq:psi} in the form of $\phi-\phi_0$, where $\phi_0$ 
is the fiducial longitude, corresponding to the moment of time when the LOS passes closest to the magnetic axis 
(Fig.~\ref{fig:geom}). Since the illumination of the emission cone can be uneven or patchy, it may be difficult to
determine the location of the fiducial longitude  based on the shape of the total intensity profile at a single frequency. 
Linearly polarised emission can provide some clues to the location of $\phi_0$: if the PA curve has an S-shape, 
as prescribed by the RVM, then, neglecting relativistic corrections, the fiducial longitude should lie  at the inflection 
point of the curve \citep[]{Radhakrishnan1969}. Unfortunately, \src's PA has a seemingly linear dependence on $\phi$ 
without an obvious inflection point \citep{Suleymanova1998,Rankin2006b,Rosen2008}. 

The location of $\phi_0$ at 430\,MHz has been constrained by DR01 and \citet{Rankin2006b} with the help of the 
cartographic analysis (i.e. mapping individual pulses to the sparks in the carousel). The authors showed that at 
the onset of a B mode, $\phi_0$ is located between two closely
spaced components of comparable amplitudes. 
As a mode ages, the relative amplitude of the trailing component diminishes and the profile acquires an asymmetric 
shape, with $\phi_0$ lying close to the trailing half-peak longitude.

Below 100\,MHz, the amplitude of the second component exhibits a similar evolution with mode age, but the components 
are farther apart from each other and have little overlap, making it easier to decompose the average profile into 
Gaussian components. As was shown by B14, the midpoint between profile components follows the $\nu^{-2}$ dispersion 
law at least between 30 and 80\,MHz. This implies that the position of the midpoint is frequency independent and is 
thus well suited for the role of $\phi_0$. However, quite surprisingly, it has been discovered that this 
frequency-independent midpoint gradually shifts  towards later spin phase, accruing an offset of 
$\delta \phi \approx 4\times10^{-3}$ by the end of a mode instance, and supposedly re-setting its position at the 
next Q-to-B transition \citep[B14,][]{Suleymanova2014}. The reason for this delay is currently unclear and may 
involve, among others, changing the illumination pattern in the emission cone or a change in emission altitude
\citep[see the discussion in B14 and][]{Suleymanova2014}.

According to Eq.~\ref{eq:theta}, the shape of phase tracks should be symmetric with respect to $\phi_0$, thus the 
phase tracks of sufficient length and quality could, in principle, provide an alternative way to measure $\phi_0$ 
and thereby offer an insight on the origin of the profile delay. For example, a changing illumination pattern 
should not affect the position of the symmetry point, while changing the emission height should result in shifting 
$\phi_0$. 

We have fitted all phase tracks from all sessions with two assumptions: (1) $\phi_0$ stays constant throughout B mode, 
fixed at its position at the midpoint between components at the beginning of the mode, and (2) that it follows the 
midpoint obtained by the two-Gaussian fit from B14. These two assumptions resulted in equally good fits with an insignificant 
difference in $N$ (Fig.~\ref{fig:phtr_fid}). Thus, we can conclude that the shape of the phase tracks did not constrain 
the location of $\phi_0$. Further improvements can be achieved with more sensitive observations and perhaps moving 
to the lower frequencies, which would expand the size of the on-pulse window.

\begin{figure*}
\centering
\includegraphics[scale=0.9]{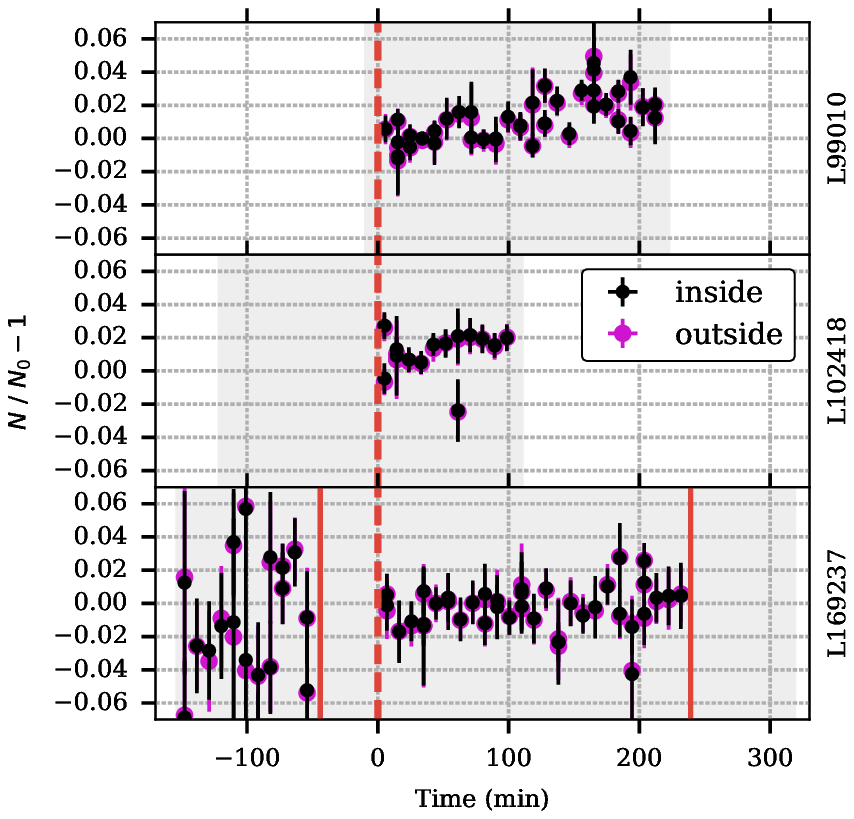}\includegraphics[scale=0.9]{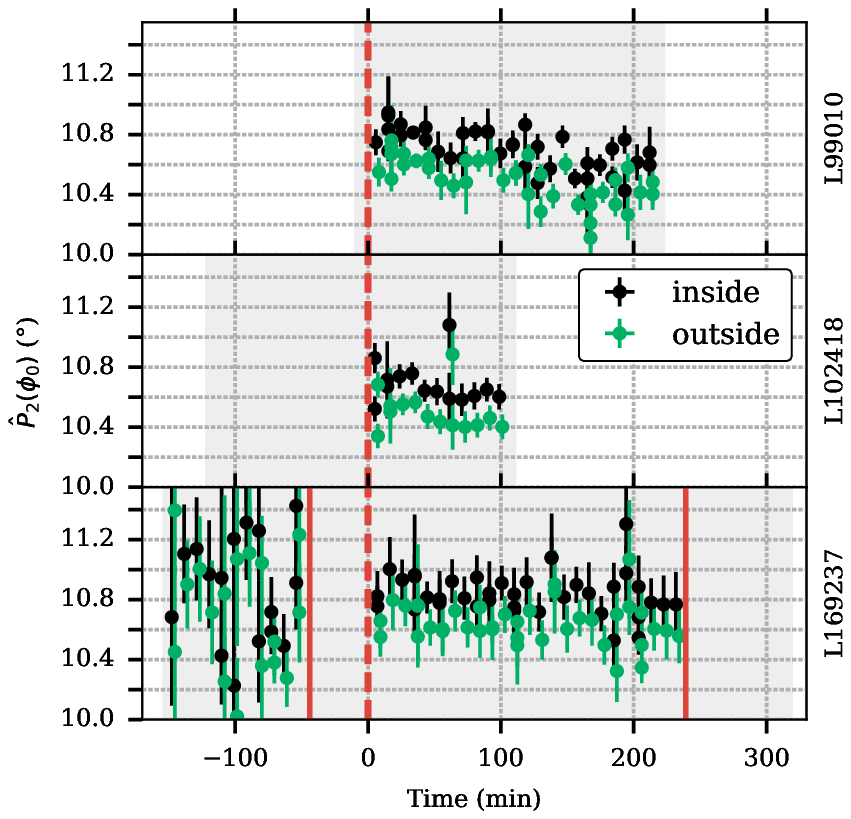}
\caption{\textit{Left:} Relative change in the number of sparks for every distinct peak in each 
of the 512-pulse LRF spectra. The fit was performed with the standard set of aliasing/geometry parameters, 
for the inside (black dots) and outside (purple dots) traverse geometries. The ``reference'' number of sparks, $N_0$, 
was taken from Fig.~\ref{fig:N_geom}. \textit{Right:} $\hptwo$ at the fiducial longitude , $\phi_0$, from Eq.~\ref{eq:P2} 
for inside (black dots) and outside (green dots) traverse geometries.}
\label{fig:p2}
\end{figure*}

\subsection{Dependence of the number of sparks on aliasing and geometry}
\label{subsec:Ngeom} 

The fitted number of sparks, $N$, depends on the assumed geometry and the degree of aliasing, $n$. In order to 
investigate this dependence, we fitted Eq.~\ref{eq:theta} to $\thph$ from a 512-pulse block with a strong narrow 
drifting feature in the LRF spectrum using the set of trial geometry angles and assumed $n$. Since the number of sparks 
may change throughout the mode (Sect.~\ref{subsec:N_vs_t}), we
label this single-block estimate of $N$ the 
''reference number of sparks'', $N_0$.

For the fit, we used the geometry solutions given by the three trial values of $\RPA$ and 13 equally spaced trial 
values of $\rhoG$, $2\degree \leq\rhoG\leq 8\degree$. In addition, we varied the aliasing order between $-2$ and $2$. 
The fits showed that $N_0$ was most sensitive to the assumed $\RPA$, then to the degree of aliasing, and only then 
on the assumed $\rhoG$ (Fig.~\ref{fig:N_geom}). Outside traverse geometries generally yielded approximately twice smaller $N_0$ than the inside traverse ones. For the standard geometry/aliasing set of parameters, $N_0=8.9$ 
and $17.6$ for the inside and outside traverse, respectively.

Interestingly, for the inside traverse geometry with $\RPA=-2.7$\,$\degree/\degree$ and $n=1$, as determined by DR01, 
our fit yields $N_0\approx19$, which is similar to their $N=20$. Thus, our results are in agreement with DR01, although 
abandoning the tertiary modulation constraint expands the range of possible $N_0$ values.

\subsection{Evolution of $N$ and $\hptwo$ throughout B mode}
\label{subsec:N_vs_t}

As was discussed in Sect.~\ref{subsec:Ngeom}, the number of sparks, $N$, depends on the specific values of 
aliasing/geometry parameters used for the fitting. The relative variation in spark number,  $N/N_0-1$, 
however, does not depend  on the exact values of the fitting parameters. Figure~\ref{fig:p2} (left) shows $N/N_0-1$
for every distinct peak on the LRF spectra in each of the 512-pulse blocks. The number of sparks at any moment under 
different aliasing/geometry assumptions can be recovered by combining $N/N_0-1$  with $N_0$ from Fig.~\ref{fig:N_geom}. 

The fitted values of $N$ jitter throughout a B mode with a characteristic magnitude of $\delta N / N_0 \lesssim 0.02$. 
This relative difference can be attributed to varying the integer number of sparks only if $N_0$ is large: 
$N_0=1/\delta N\gtrsim1/0.02=50$. This can be achieved only by assuming a very high degree of aliasing. However, any 
conclusions here should be drawn with caution, since the nature of the drift frequency jitter (or splitting, within a 
single pulse block) is unclear and $\thph$ can be contaminated by phase jumps or phase modulation.

Apart from this erratic behaviour, $N/N_0$ increases by $\approx 0.02$ on a timescale of mode duration for the first 
two observations (L99010 and L102418). Correspondingly, $\hptwo$ for the first two sessions decreases by about 
$0.2\degree$ (Fig.~\ref{fig:p2}, right).

It is interesting to compare these results with those obtained by \citet{Backus2011}. The authors computed HRF spectra 
on an unfolded time series and examined the dependence of modulation intensity on the harmonic number for the  
primary modulation feature. $\hptwo$ was calculated from the harmonic number corresponding to the peak of the 
modulation intensity, $\hptwo = 360\degree/N_\mathrm{harm}$. The $\hptwo$ obtained was about $11\degree$ at the beginning
of the B mode and slowly decreased to about $10\degree$ in a seemingly exponential manner. The authors did not give 
the measurement error, but the points in their plot are scattered by about $0.5\degree$. The authors considered 
the on-pulse window to be small enough to neglect the longitudinal variation of $\hptwo$.

For the later stages of a B mode, our fit suggests a larger $\hptwo$, of about $10.6\degree$, even for the lowest 
possible value of $\hptwo$ at the fiducial longitude. The values of $\hptwo$ in our fit depend on $\RPA$, with 
$\hptwo$ for $\RPA=-2.4\degree/\degree$ being larger by $0.2\degree$ and smaller by the same amount for 
$\RPA=-3.4\degree/\degree$. Even if we obtain $\hptwo$ using exactly the same geometry as in DR01, it is still larger 
than in \citet{Backus2011}, and the reasons for this are currently unclear.

In our observations, the gradual variation of $\hptwo$ throughout the B mode is about five times smaller
than reported by 
\citet{Backus2011}. A possible explanation for this would be the following: \citet{Backus2011} investigated HRF spectra, 
thus effectively probing $\hptwo$ at all longitudes within the on-pulse window.  The size of the on-pulse window at 
their frequencies is $\approx10\degree$, and $\hptwo$ changes by about $1\degree$ between the centre of the window and 
its edges (Fig.~\ref{fig:phtr}). The relative contribution of $\hptwo(\phi)$ at each longitude may be substantially 
influenced by the average intensity of emission at that longitude. At the radio frequency of \citet[][325\,MHz]{Backus2011}, 
the intensity of the emission at the longitudes of trailing component undergoes substantial changes throughout the B mode: at the
mode onset, the trailing component is approximately of the same height as the leading component, and at 
the end of the B mode, the amplitude of the trailing component is only about 20\% of the leading component amplitude 
\citep{Rankin2006b}.
Thus, at the beginning of the B mode, the longitude window of the trailing component, where $\hptwo(\phi)>\hptwo(\phi_0)$, 
is illuminated better than at the end of the mode, and this might cause observed increase in the derived  
$\hptwo$. The authors investigated the 
influence of profile shape evolution on the measured $\hptwo$. However, they explicitly simulated the intrinsic $\hptwo$ 
being constant within the on-pulse window.

The observed small changes in $\hptwo$ may be due to the evolution of the linear polarisation.
According to \citet{Suleymanova1998} and \citet{Rankin2006b}, at frequencies of 100--300\,MHz,  
the linearly polarised emission in both profile components consists of two orthogonal 
modes of unequal strength.  
It has been shown that these two polarisation modes correspond to two sets of sparks in the carousel, shifted 
in azimuth with respect to each other (DR01). Moreover, observations at 64 and 327\,MHz revealed that 
the fractional linear polarisation at the peak of the leading component 
evolves substantially throughout B mode \citep{Suleymanova2009}, perhaps indicating the 
change in the relative strength of the polarisation modes. If the relative intensity of the two 
sets of sparks changes at LBA frequencies, it can add a time-dependent bias to the phase tracks, 
which may result in the observed variation in $\hptwo$. Another pulsar, for which 
 the incoherent superposition of the two orthogonal polarisation modes led to biased measurements of 
$\hptwo$ is PSR~B0809+74 \citep{Rankin2005,Rankin2014}.

 \begin{figure}
\centering
\includegraphics[scale=0.9]{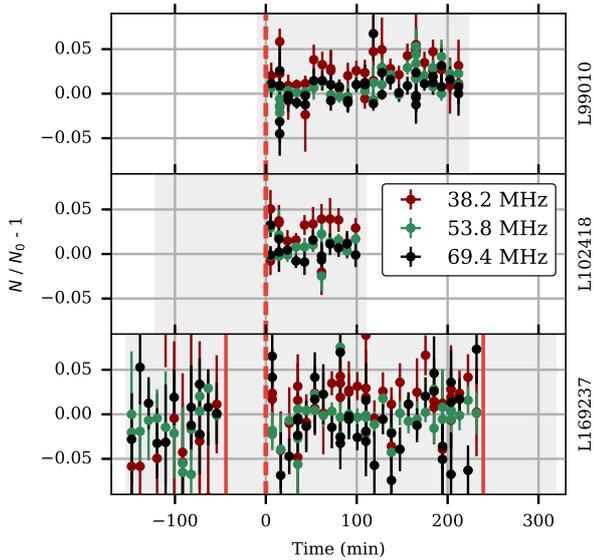}
\caption{Relative change in the number of sparks for the three frequencies within the LBA band with respect to the 
reference number of sparks $N_0$ from Fig.~\ref{fig:N_geom}. The fit at the lowest frequency yields slightly but consistently 
larger $N$.} 
\label{fig:Nfr}
\end{figure}

\begin{figure*}[t]
 \centering
 \includegraphics[scale=0.9]{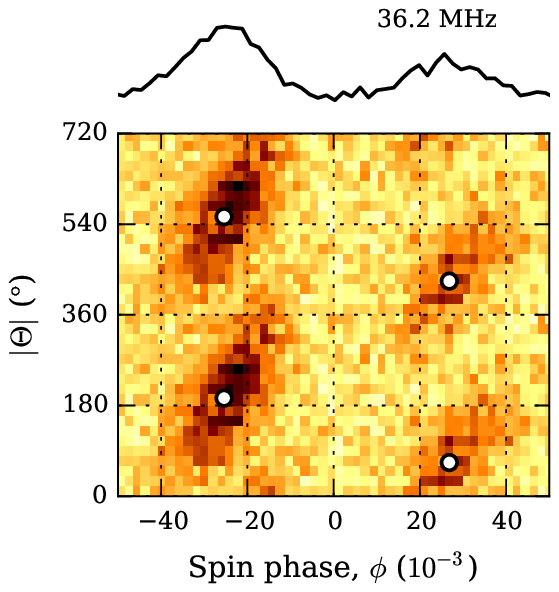}\includegraphics[scale=0.9]{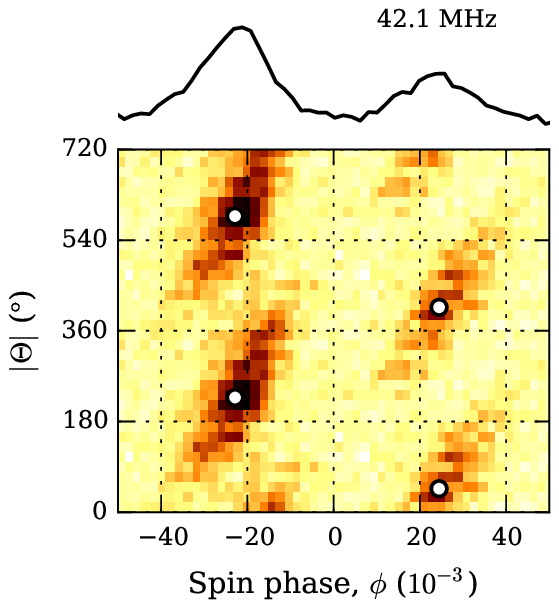}\includegraphics[scale=0.9]{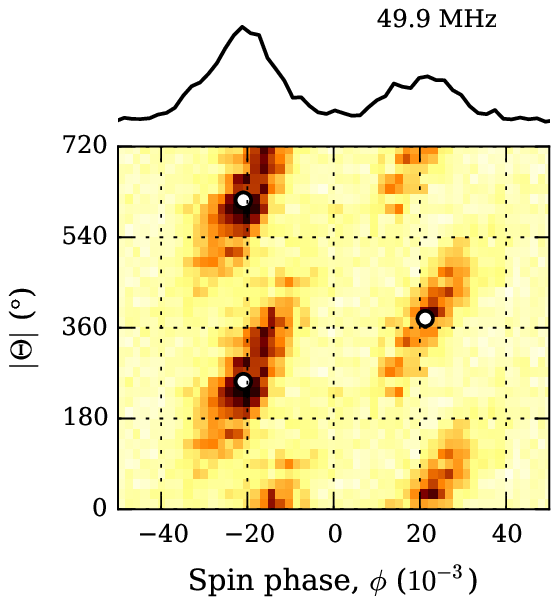}
 \includegraphics[scale=0.9]{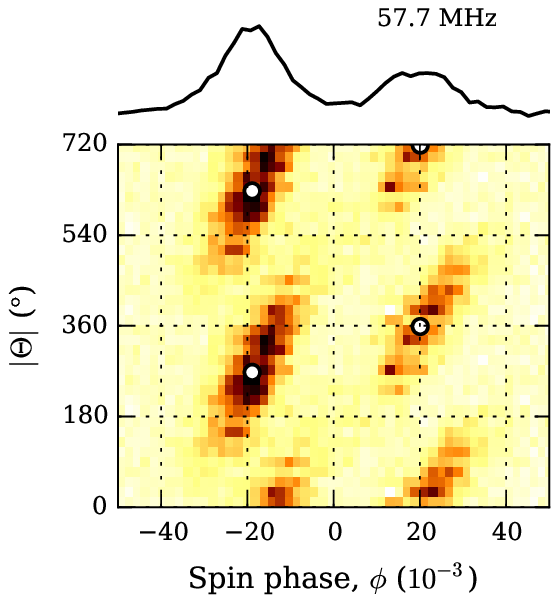}\includegraphics[scale=0.9]{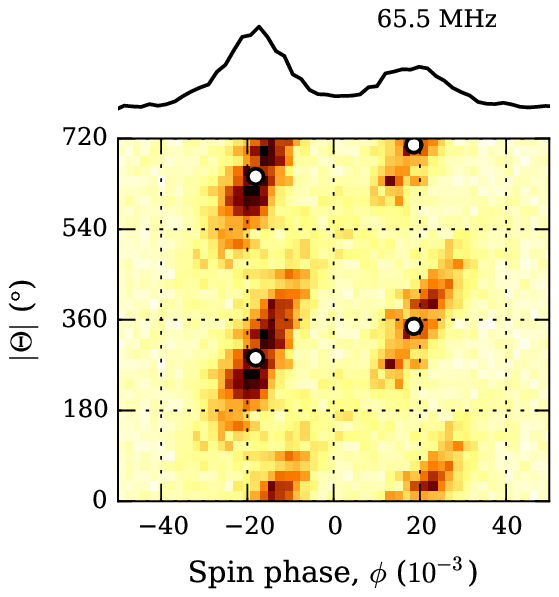}\includegraphics[scale=0.9]{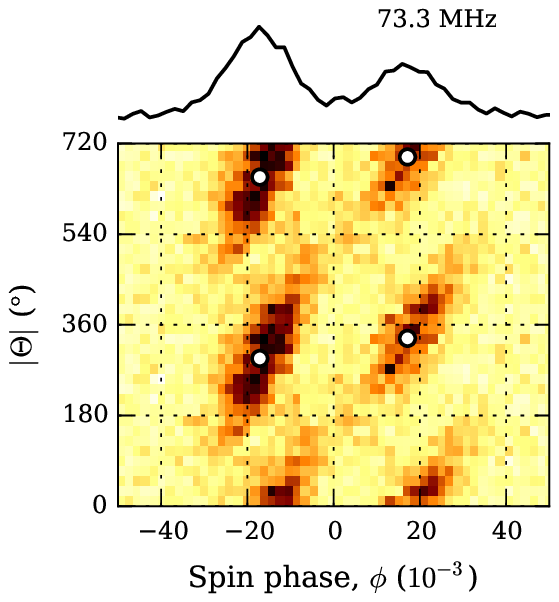}
 \caption{Example of a 512-pulse stack folded modulo $\hpthr$  for six radio frequencies within the LBA band. 
 The modfold is repeated twice along the Y-axis to mitigate wrapping, and the corresponding average profile is plotted on 
 top. White circles mark the centres of driftbands obtained from fitting 2D tilted Gaussians. As the radio 
 frequency increases, the driftbands move towards each other both in $\phi$ (reflecting the behaviour of the average 
 profile components) and  in $\Theta$.}
 \label{fig:driftbands}
\end{figure*}
    
\subsection{Number of sparks versus radio frequency}

We fitted $N$ and $\theta_0$ to three 16-MHz subbands in the middle of the LBA band, where the S/N was highest (at 38.2, 53.8, and 69.4\,MHz). The fit was performed with the standard geometry/aliasing parameters. For all pulse blocks and 
every pair of frequencies, $|N(\nu_1)-N(\nu_2)|$ divided by the the quadrature sum of the corresponding uncertainties was 
lower than 3. However, there is a slight tendency for the lowest frequency to yield a higher value of $N$ (Fig.~\ref{fig:Nfr}, left). 

A frequency-dependent number of sparks would be hard to explain with the carousel model, but a deeper investigation is 
needed  to confirm or refute this tendency: for example with calibrated polarisation to resolve the orthogonal modes and  
with better sensitivity at frequencies below 40\,MHz. 

\begin{figure*}
   \centering
 \includegraphics[scale=1.0]{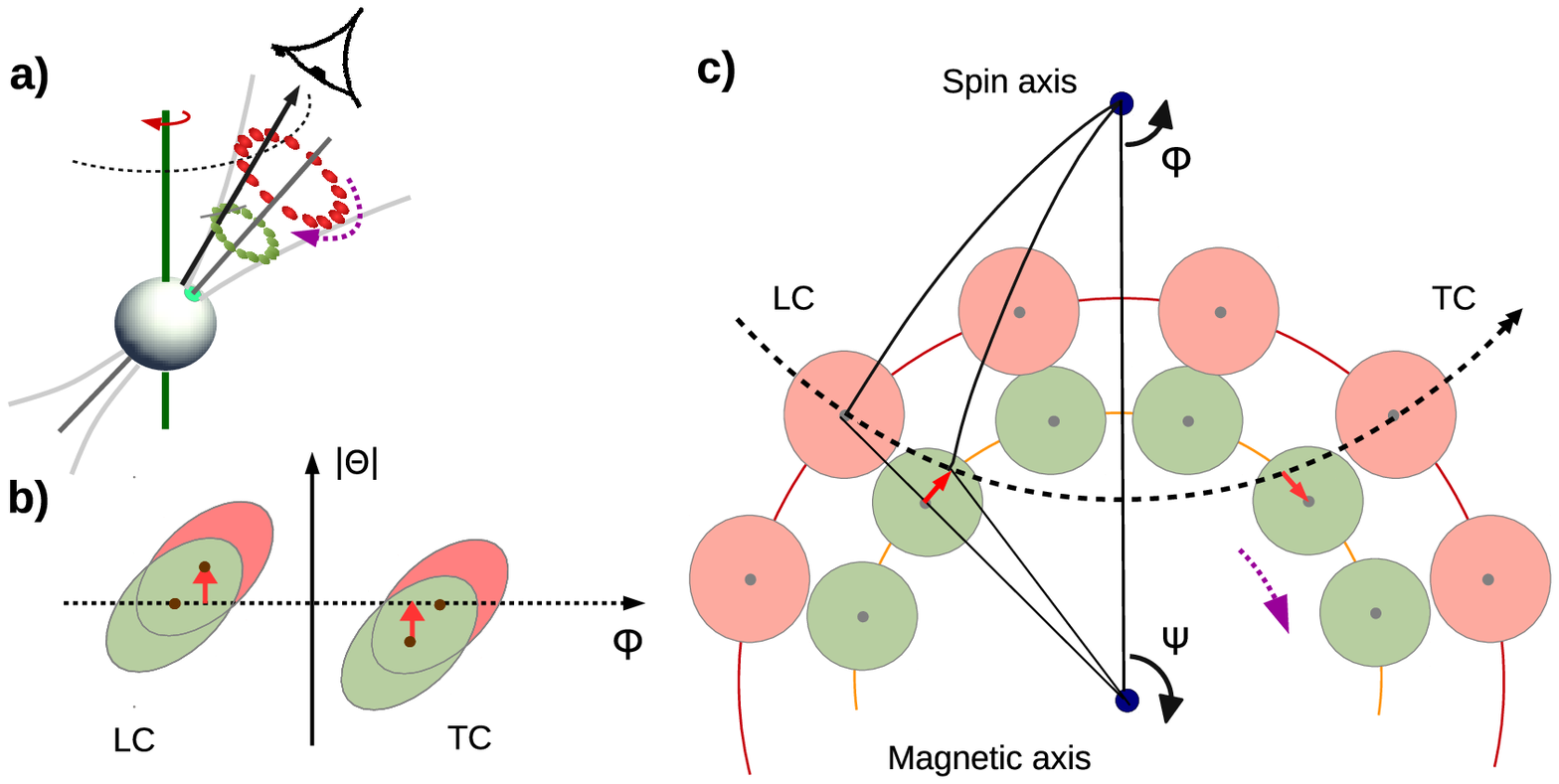}
 \caption{Cartoon model of the drift track evolution: \textit{(a)} general view on the system (inside traverse geometry,
 pulsar is rotating clockwise), 
 similar to Fig.~\ref{fig:geom}, but with individual sparks at two different radio frequencies, 
 $\nulo$ (red) and $\nuhi$ (green). According to the RFM, emission at $\nulo$ comes from higher altitudes, where the opening 
 angle of the dipole field lines is larger. The modfolds at these two frequencies are shown in \textit{(b)}. The carousel 
 configuration corresponding to the moment marked with the dotted horizontal line in \textit{(b)} is shown in \textit{(c)}. 
 The violet arrows in (a) and (c) mark the direction of the carousel rotation (see text for more details). 
 The black arrows show increasing longitude, $\phi$, and the magnetic azimuth, $\psi$, both measured from the fiducial longitude.
 At this moment, the LOS sweeps 
 through the spark centres at $\nulo$ and misses the spark centres at $\nuhi$. For the leading component (LC), the 
 rotating carousel needs more time to bring the spark centre to the LOS. The drift phase $\Theta$ is directly proportional 
 to time (Eq.~\ref{eq:Theta}), thus  $|\TLC(\nuhi)|>|\TLC(\nulo)|$. The  spherical triangles defined by the spin axis, 
 magnetic axis, and the point where the LOS passes through the emission cone (black lines) are identical to the triangle in 
 the inset of Fig.~\ref{fig:geom}. 
 }
 \label{fig:Theta_cartoon}
    \end{figure*}

\section{$\hpthr$ folds in B mode}
\label{sec:fdphd}

In the schematic depiction of drifting subpulses in Fig.~\ref{fig:P2P3}, the individual drift tracks are clearly 
visible because of the relatively large $\hpthr$ and the uniformly high S/N of the individual pulses. For \src, 
$\hpthr$ is comparable to $\pone$ and the amplitudes of the pulses can vary substantially, preventing the shape 
of the drift tracks from being evident in the pulse stacks alone (Fig.~\ref{fig:ps}). In order to uncover the 
shape of the drift tracks, it is customary to fold the pulse stacks modulo $\hpthr$, assigning each period in the pulse stack 
$s(\phi,t)$ a ``drift phase'':
\begin{equation}
\Theta = \frac{t\pone}{\hpthr}\times360\degree, 
\end{equation}
where $t$ is an integer pulse number, the same as in  Eq.~\ref{eq:sdrift}. Such folds are called ``modfolds'' or 
``driftbands'' in the literature \citep{Backus2011, Hassall2013}. 

At higher radio frequencies, the \src\  modfolds consist of a single drift track spanning the entire on-pulse 
window \citep[350\,MHz, DR01,][]{Backus2011}. Around 100\,MHz, the drift track is still continuous, but wraps twice 
per $\hpthr$ (DR01). Finally, at the lowest radio frequencies,  two separate drift tracks are visible in the 
longitude regions of two distinct profile components \citep[35\,MHz,][]{Asgekar2001}. 

Figure~\ref{fig:driftbands} shows an example of the modfold for a single pulse stack, folded with 
the corresponding $\hpthr<0$ from Sect.~\ref{sec:Bsubp}. Unlike DR01, we did not subtract any ``aperiodically 
fluctuating baseline`` from the modfolds, and for most of the pulse stacks, the drift tracks had a well-defined 
shape with little baseline signal between them. In a few cases, however, when folding the pulse stacks with 
multiple peaks in the LRF spectrum, the drift tracks appeared to be blurred. We note that the drift tracks in 
Fig.~\ref{fig:driftbands} are tilted to the right (pulses drift towards the trailing edge of on-pulse window), 
whereas in the literature the drift tracks are tilted to the left (pulses drift towards the leading edge). This 
discrepancy stems from difference in the period, $\pthr$, used for folding -- after DR01, all modfolds in the 
literature were performed with $\pthr$ defined by the relation
\begin{equation}
\dfrac{\pone}{\pthr} = n+\dfrac{\pone}{\hpthr},
\label{eq:P3}
\end{equation}
where the degree of aliasing was taken to be $n=1$. Drift tracks look exactly the same for any  $|n|$, but 
for all $n\leq0$ the tracks are tilted to the right and for all $n>0$ they are tilted to the left.

The white circles in Fig.~\ref{fig:driftbands} mark the centres of the drift tracks obtained from fitting 2D 
tilted Gaussians. We label the coordinates of the track centres ($\phi_\mathrm{C}$, $\TC$), adding L for 
the leading component and T for the trailing component when needed. Both $\phi_\mathrm{C}$ and $\Theta_\mathrm{C}$ 
appear to evolve with radio frequency, approaching each other at higher $\nu$.  This creates the apparent 
diagonal movement of the drift tracks, which continues beyond the LBA band; around 100\,MHz, the drift tracks fuse
at the inner corners and continue to merge, eventually forming a single drift track at 350\,MHz. 
For the modfolds folded 
with $n>0,$ the direction of vertical movement is reversed for both drift tracks.
The $\phi_\mathrm{C}$, 
obviously, corresponds to the position of the average profile components, and 
the decreasing $|\phi_\mathrm{C}-\phi_0|$ reflects the merging components at higher frequencies.
The longitude spacing
between driftband centres does not correspond to $\hptwo$, as the intensity within driftbands is set by the shape of the  
average profile.

A similar frequency-dependent $\Theta$ was found by \citet{Hassall2013} in the modfolfds of PSR~B0809+74. Unlike 
\src, the $\TLC$ for the pulsar B0809+74 was frequency independent and the $\Theta_\mathrm{TC}$ was moving towards 
$\Theta_\mathrm{LC}$ with increasing $\nu$. The same behaviour was exhibited by $\phi_\mathrm{C}$ : the position of 
the leading component was frequency independent while the trailing component was moving towards it with increasing 
radio frequency. \citet{Hassall2013} did not give an explanation to the observed $\Theta_\mathrm{C}$ behaviour. 
However, \citet{Rankin2014} later suggested that it might be explained by the frequency-dependent phase 
delay of the rotating carousel with respect to the LOS.

Below we show that for \src,\ the observed $\TC(\nu)$ can be quantitatively explained within the carousel model. 
Without loss of generality, we consider a simplified model with simultaneous 
observations at only two radio frequencies, $\nuhi>\nulo$. For simplicity, we assume that during the chosen traverse, the 
LOS cuts through the spark centres at $\nulo$. At $\nuhi$, the same LOS sweep will miss the spark centres, 
causing the offset between drift tracks at different frequencies. Figure~\ref{fig:Theta_cartoon} shows a cartoon model 
of this effect for the inside traverse geometry and $n=0$ modfolds.  For the inside traverse geometry solution, the 
pulsar rotates clockwise (Fig.~\ref{fig:traverse}), thus the LOS sweeps from left to right on the image. Folding 
pulse stacks with $\pthr$, determined by $n\leq0$ (Eq.~\ref{eq:P3}) implies that from the observer's point of view, the 
carousel rotates clockwise (pulses drift towards the trailing edge of profile). Altogether, this means that at 
$\nuhi$ LOS will cut through the leading part of the spark for the leading component and that the carousel will need 
extra time to rotate to bring the spark centre to the LOS (Fig.~\ref{fig:Theta_cartoon}). In the modfold plots, time increases 
upwards, thus $|\TLC(\nuhi)|>|\TLC(\nulo)|$ and $|\TTC(\nuhi)|<|\TTC(\nulo)|$. Folding pulse stacks with  $\pthr$, 
determined by $n\leq0$, reverses the assumed direction of the
carousel rotation and the direction of the $\TC$ shift, exactly 
as observed. Finally, the same reasoning applies for the outside traverse geometry solution.

\begin{figure}
\centering
\includegraphics[scale=0.9]{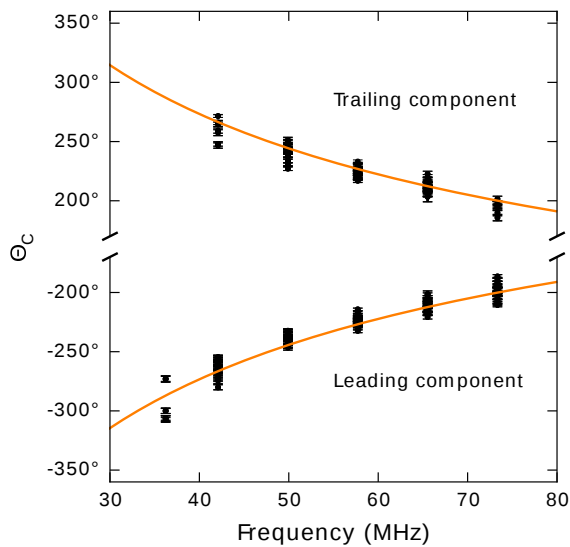}
\caption{Black points show the drift phase $\TC$ vs. radio frequency for the leading and 
trailing drift tracks for all pulse stacks of the session L99010, where the fitted 2D Gaussians had an $\mathrm{S/N}>7$. 
The orange lines mark the expected phase delay of the rotating carousel calculated according to Eq.~\ref{eq:Theta}.}
\label{fig:phi3_nu}
\end{figure}

Quantitatively, the drift-phase delay between two frequencies can be expressed as\label{eq:Theta_prep}
\begin{multline}
 \Theta_\mathrm{C}(\nulo) - \Theta_\mathrm{C}(\nuhi) =  -N\,\mathrm{sgn}\,\beta\times[\psi_\mathrm{C}(\nulo)-\psi_\mathrm{C}(\nuhi)] +  \\ + [\phi_\mathrm{C}(\nulo)-\phi_\mathrm{C}(\nuhi)]\times\frac{\pone}{\pthr}. 
\end{multline}
Here the first term represents the scaled difference in magnetic azimuth angles calculated with Eq.~\ref{eq:psi} at 
$\phi_\mathrm{C}(\nuhi)$ and $\phi_\mathrm{C}(\nulo)$. The second term accounts for the carousel rotation during the time 
it takes the LOS to travel between $\phi_\mathrm{C}(\nuhi)$ and $\phi_\mathrm{C}(\nulo)$. 

\citet{Bilous2014} have shown that within the LBA band, $\phi_\mathrm{C}$ can be approximated with the following relation:
\begin{equation}
\label{eq:phiC}
\phi_\mathrm{TC/LC}(\nu) = \phi_0 \pm 69\degr.84\times\left[\frac{\nu}{1\,\mathrm{MHz}}\right]^{-0.567}.
\end{equation}
Extrapolating this relation to infinite frequency yields $\phi_\mathrm{C}(\nu=\infty)=\phi_0$ for both profile components. 
This can be used to rewrite Eq.~\ref{eq:Theta_prep} with respect to $\nuhi = \infty$:
\begin{equation}
\label{eq:Theta}
 \Theta_\mathrm{C}(\nu) =  -N\,\mathrm{sgn}\,\beta\times\psi(\phi_\mathrm{C}) +(\phi_\mathrm{C}-\phi_0)\frac{\pone}{\pthr} + \Theta_0, 
\end{equation}
where $\Theta_0= \Theta_\mathrm{C}(\nu=\infty)$.

Figure~\ref{fig:phi3_nu} shows such $\Theta_\mathrm{C}(\nu)$ calculated for the standard geometry/aliasing parameters, for
the reference 
number of sparks, $N_0$, and $\phi_\mathrm{C}$ from Eq.~\ref{eq:phiC}. The phase at the fiducial longitude $\Theta_0$ 
is set to 0 in this figure. Measurements of $\Theta_\mathrm{C}$ from the Gaussian fit to drift tracks are overplotted for all phase tracks from 
the observing session L99010 for which the S/N of the drifting feature in the LRF spectra was higher than six
(modfolds from the other two sessions exhibited the same $\Theta_\mathrm{C}(\nu)$ dependence).  The modfolds were folded with $\pthr$
determined by the corresponding distinct $\hpthr$ from the peaks on the LRF spectra and $n=0$. 
Since each individual modfold was produced with sightly different $\hpthr$, the 
fiducial longitude phase $\Theta_0$ is essentially random, and thus should be subtracted from $\TC$. To do this, we investigated 
$0.5(\TLC+\TTC)$ for the drift tracks for which the amplitudes of fitted 2D Gaussians had an $\mathrm{S/N}>7$. For most of these cases, 
this mean value of $\TC$ did not have a dependence on $\nu$, thus we took it as $\Theta_0$. Since we set the point of 
symmetry manually, the fixed and moving $\phi_0$ hypotheses from Sect.~\ref{subsec:phi0} will both result in an identical $\TC$ behaviour.

The agreement between the measurements and the prediction from the rotating carousel model is very good, although the scatter 
of points is larger than the formal fit errors. Several factors may contribute to the additional scatter: $N$ may slightly 
vary throughout the B mode (Fig.~\ref{fig:P3_vs_t}), or the shape of the driftbands may deviate from simple Gaussian due 
to uncertainties in $\hpthr$, the small number of $\Theta$ bins (18), and  the small number of pulses ($\sim15$) that are averaged 
in a single $\Theta$ bin.

\section{Summary and conclusions}
\label{sec:summ}

\src, with its two discrete metastable emission modes and the slow variation of the emission properties in (at least) one of 
the modes, provides a great opportunity to investigate the non-stationary magnetospheric configurations and to untangle the complex 
interplay between various physical processes in a pulsar magnetosphere. However, with many new facts that 
come to light as a result of improved observational capabilities in X-rays and at the lowest radio frequencies, 
it becomes increasingly important to keep track of the 
assumptions made to interpret previous observations.

The very basic question of the \src\  geometry is still not answered fully. In Sect.~\ref{sec:geom} we have successfully reproduced 
the original geometry derivation from DR01 using the broadband LOFAR observations at 25--80\,MHz. DR01 argued 
that \src\ is an almost aligned rotator, with an inclination angle $\alpha$ within $12\degree$--$16\degree$, and the LOS passing 
between the spin and magnetic axes, yielding an impact angle $\beta$ of about $4\degree$--$6\degree$ (the angles are given 
in EW01 notation, which is different from the one used by DR01). We must stress, however, that the exact values of the geometry angles 
depend heavily on the assumed opening angle of the emission cone at 1\,GHz, $\rhoG$, namely $\alpha\approx 3 \rhoG$ (or 
$180\degree$--$\alpha\approx 3 \rhoG)$, and $\beta\approx\rhoG$. The distribution of $\rhoG$ from the currently largest such 
work (MD99) yields $\alpha$ (or $180\degree-\alpha$) of $5\degree$--$25\degree$, leaving it somewhat less strongly constrained than in DR01.
In principle, the current geometry derivation scheme allows for almost any values of $\alpha$, since there are no theoretical 
constraints on $\rhoG$, and the 37-source sample of $\rhoG$ from MD99 is still rather small. 

The weaker constraints on geometry angles have to be taken into account  when the unexpectedly large fractional amplitude
of the thermal X-ray emission from the polar caps of \src\ is
explained \citep{Hermsen2013, Mereghetti2016, Mereghetti2017}. Another interesting 
constraint on the \src\  geometry may come from directly modelling the X-ray light curve.

Unlike DR01, we chose not to exclude the possibility of the LOS passing equatorwards from the magnetic axis (the 
so-called ''outside traverse``).
The decision of DR01 was based on the brief detection of two sidelobes around the main drifting feature in the LRF B-mode spectra.
These sidelobes were interpreted as the amplitude modulation caused by a rotating carousel of $N=20$ sparks. Despite dedicated 
searches and a relatively large volume of B-mode observations, we did not detect any sign of such amplitude modulation, thus concluding 
that it must be quite rare, or a narrowband phenomenon. Considering the 
objection raised in \citet{Rosen2008}, 
we decided to discard the $N=20$ constraint in all our subsequent analysis. This resulted in an unconstrained aliasing order and multiple
possible values for the number of sparks in the carousel.

Similar to the previous works, we did not find any periodic subpulse modulation in the Q mode and confirm the gradual frequency-independent 
evolution of drift period $\hpthr$ (and thus the carousel circulation time) in B mode. We must note that with the unknown aliasing order, 
the relative slow-down of the carousel is unconstrained (the carousel may even be spinning up), which should be taken into 
account while modelling the conditions in the polar gap \citep{Backus2011,Szary2015}.

Very many of our 512-pulse LRF spectra showed multiple peaks, corresponding to distinct values of $\hpthr$. Observing two 
values of $\hpthr$ simultaneously would serve as a strong corroboration of the drifting subpulses surface oscillation model 
\citep{Rosen2008}. Although our ability to resolve true temporal simultaneity was limited by the frequency resolution of the 
discrete Fourier transform, we were able to show that for all 512-pulse LRF spectra with relatively widely separated peaks,
only  one peak was found in each half of a pulse sequence, indicating that $\hpthr$ may change on a timescale as short as $256\pone$.
It is currently unclear whether the observed multiple peaks are related to a true change in $\hpthr$ (e.g. some of them 
may be caused by jumps in drift phase), but if genuine, they would point to few-minute $\hpthr$ variations with a  
magnitude only a few times smaller than the overall $\hpthr$ change over the mode duration.

We obtained the number of sparks under different aliasing/geometry assumptions by fitting the drift phase tracks with 
the analytical prescription of the carousel model (ES01). Multiple peaks on a single LRF spectrum yielded an integer variation 
in the number of sparks only if the initial number of sparks was large ($N\gtrsim 50$). We find a weak evidence of $N$ being 
larger at lower radio frequency, although more sensitive/broadband observations are needed to confirm this. 

Phase tracks were also used to estimate the longitudinal spacing between sparks, $\hptwo$. Unlike \citet{Backus2010}, we did 
not detect a large ($\approx1\degree$) gradual change of $\hptwo$ in any of our B-mode observations. We speculate that the 
apparent $\hptwo$ evolution of \citeauthor{Backus2010} may in fact be caused by $\hptwo$ variation within the on-pulse window, 
which is illuminated differently throughout the B mode. Our data suggest a much smaller ($0.2\degree$) variation of $\hptwo$. 
 
This variation may stem from the bias to phase track caused by incoherent mixing of two linear polarisation modes of variable relative strength
\citep{Rankin2006b}, each mode corresponding to a separate set of sparks shifted in azimuth with respect to each other (DR01). If real, the 
variations of $\hptwo$ may be linked to the changing plasma density in the magnetosphere through the motion of the visibility 
point \citep{Yuen2016}.

Finally, we described the frequency-dependent drift phase delay, similar to the delay reported by \citet{Hassall2013} for 
PSR B0809+74. We show that this delay can be quantitatively modelled within the rotating carousel model in the RFM convention.
It would be very interesting to explain this effect in the framework of other drifting subpulse models 
\citep[e.g. surface oscillations from][]{Rosen2008} or non-RFM pulsar emission models \citep{Dyks2015}. 
A similar quantitative explanation can be also applied to PSR B0809+74 (Bilous et al., in prep).

\begin{acknowledgements}
AVB thanks the referee, Joanna Rankin (UVM), as well as Vlad Kondratiev (ASTRON), and Tim Pennucci (ELTE) for the valuable comments 
on the manuscript.
This paper is based (in part) on data obtained with the International LOFAR Telescope (ILT). LOFAR \citep{vanHaarlem2013}
is the Low Frequency Array designed and constructed by ASTRON. It has facilities in several countries, which are owned by 
various parties (each with their own funding sources), and which are collectively operated by the ILT foundation under 
a joint scientific policy.
\end{acknowledgements}

\bibliographystyle{aa} 
\bibliography{0943_bibliography}

\end{document}